\documentclass[lettersize,journal]{IEEEtran}
\usepackage{cite}
\usepackage{amsmath,amssymb,amsfonts}
\usepackage{algorithmic}
\usepackage{graphicx}
\usepackage{textcomp}
\usepackage{xcolor}
\usepackage{algorithm}
\usepackage{array}
\usepackage{stfloats}
\usepackage{url}
\usepackage{verbatim}
\usepackage{makecell}
\usepackage{color} 
\usepackage{xpatch}
\usepackage{setspace}
\usepackage{cuted}
\usepackage{subfigure}

\def\BibTeX{{\rm B\kern-.05em{\sc i\kern-.025em b}\kern-.08em
		T\kern-.1667em\lower.7ex\hbox{E}\kern-.125emX}}
\usepackage{balance}
\makeatletter
\def\changeBibColor#1{%
	\in@{#1}{11775961,11185309,10901277,Chi2026CDMAMP,Li2026AFDMTimeScaling,9349154}
	\ifin@\color{black}\else\normalcolor\fi	
}
\xpatchcmd\@bibitem
{\item}
{\changeBibColor{#1}\item}
{}{\fail}
\xpatchcmd\@lbibitem
{\item}
{\changeBibColor{#2}\item}
{}{\fail}
\makeatother
\begin{document}
	\title{Chirp Parameter Optimization and Distributed Detection for Cooperative RSMA-AFDM Systems}
	\author{Qingyu~Li, Guanghui~Liu, \textit{Senior Member, IEEE}, Yusha~Liu, Fuchen~Xu, Chengxiang~Liu, Hongjun~Liu, and Liaoyuan~Zeng
		\thanks{This work was supported in part by the Natural Science Foundation of China under Grant 62671119; in part by Hainan Province Science and Technology Plan Projects under Grant ZDYF2025(LALH)001, Grant ZDYF2025(LALH)003, and Grant ZDYF2026LALH001; and in part by the Natural Science Foundation of Sichuan Province under Grant 2026NSFSC0399. (\textit{Corresponding author: Guanghui Liu}).} 
		\thanks{Qingyu Li, Guanghui Liu, Yusha Liu, Fuchen Xu, Chengxiang Liu, Hongjun Liu, Liaoyuan Zeng are with the School of Information and Communication Engineering, University of Electronic Science and Technology of China, Chengdu 611731, China (e-mail: qingyu.li@std.uestc.edu.cn; guanghuiliu@uestc.edu.cn; yusha.liu@uestc.edu.cn; fuchenxu@std.uestc.edu.cn; cxliu@std.uestc.edu.cn; hjliu@std.uestc.edu.cn; lyzeng@uestc.edu.cn).
		}%
	}
	
	\markboth{IEEE TRANSACTIONS ON WIRELESS COMMUNICATIONS,~Vol.~XX, No.~XX, XX~XXXX}%
	{}
	
	\maketitle
	
	\begin{abstract}
		Affine frequency division multiplexing (AFDM) exhibits strong Doppler robustness and an efficient representation of doubly selective channels, but its signal-spreading property complicates conventional time--frequency multiple access. To address this issue, we develop a cooperative rate-splitting multiple access (RSMA)-AFDM framework. \textcolor{black}{A conditional pairwise error probability (PEP)
		expression is derived for cooperative common-stream detection, with
		the common-stream precoder explicitly incorporated into the effective
		distance. The resulting expression reveals that user-dependent chirp
		parameters influence the PEP through the DAFT-domain equivalent
		channels and the precoder constructed from them. Based on this
		\textcolor{black}{relationship, a PEP-guided cooperative effective-distance optimization is formulated for a general
		multiuser setting. The chirp parameters
		are selected by minimizing the channel-averaged common-stream
		PEP.}} \textcolor{black}{With this design,} two distributed expectation propagation (EP)-based common-stream detectors are proposed. The first combines local EP estimates and cooperative information through maximum-ratio decision fusion. \textcolor{black}{The second performs belief consensus by exchanging and aggregating external Gaussian likelihood natural parameters while iteratively refining a single shared EP site.} Simulation results confirm that the proposed chirp-parameter design and cooperative EP detectors significantly improve the detection performance and reliability of cooperative RSMA-AFDM systems.
	\end{abstract}
	
	\begin{IEEEkeywords}
		Affine frequency division multiplexing (AFDM), rate splitting multiple access (RSMA), distributed cooperative detection, belief consensus, expectation propagation (EP), chirp parameter optimization, doubly selective channels.
	\end{IEEEkeywords}

	\section{Introduction}
	\IEEEPARstart{N}{ext}-generation mobile communication systems are expected to support diverse services and massive connectivity, which is particularly challenging in high-mobility scenarios such as non-terrestrial networks (NTN) and intelligent transportation systems \cite{10845880}. High terminal velocities induce large Doppler shifts, destroying the orthogonality of orthogonal frequency division multiplexing (OFDM) \cite{9724198,9285313} and increasing channel-estimation overhead. Orthogonal time frequency space (OTFS) \cite{22} represents high-mobility channels in the delay-Doppler (DD) domain \cite{9404861,9720136}. However, practical fractional Doppler shifts cause energy leakage, resulting in less sparse DD-domain channels and considerable estimation overhead \cite{33,44,9508932}.
	
	Affine frequency division multiplexing (AFDM) \cite{99} has recently been developed to modulate signals in the discrete affine Fourier transform (DAFT) domain. Unlike OFDM with fixed-frequency subcarriers, AFDM employs chirp subcarriers whose instantaneous frequencies vary linearly with time \cite{10975107}. Since chirp signals preserve their structure under Doppler shifts, AFDM exhibits inherent Doppler robustness \cite{99,10557524,10845819}. \textcolor{black}{Its flexible chirp parameters also enable AFDM to adapt to wideband communication systems \cite{Li2026AFDMTimeScaling}.} Moreover, AFDM achieves full channel diversity using one-dimensional signaling, thereby reducing transmission latency compared with block-based waveforms such as OTFS \cite{11003079,10769778}. \textcolor{black}{Its equivalent channel also remains highly sparse under fractional Doppler shifts \cite{Liu2026AFDMPAPR}, leading to lower channel-estimation overhead and low complexity detection \cite{11775961}.}
	
	However, conventional orthogonal multiple access (OMA) in the time-frequency (TF) domain cannot be directly applied to transform-domain waveforms such as affine frequency division multiplexing (AFDM) and orthogonal time frequency space (OTFS), because each symbol is spread over the entire TF domain and signals from different users overlap. A point-constrained inverse symplectic finite Fourier transform (ISFFT) scheme was proposed in \cite{8515088} to confine each user’s symbols to a limited TF region. Although this enables user separation, it sacrifices part of the inherent channel-diversity gain. \textcolor{black}{In addition, DAFT-spread AFDMA schemes are investigated for both uplink and downlink transmissions in \cite{11185309} and \cite{10901277}, respectively, demonstrating performance advantages over conventional orthogonal AFDMA (O-AFDMA).}
	
Non-orthogonal multiple access (NOMA), including code-domain and power-domain schemes, has been investigated for transform-domain waveforms. Integrating OTFS with sparse code multiple access (SCMA) achieves high asymptotic diversity gain \cite{9411900}, while cross-domain iterative detection and decoding \cite{10022044} and DL-based methods \cite{10711268} have been developed to mitigate error propagation. For power-domain NOMA, its integration with OTFS improves error-rate and capacity performance over conventional NOMA-OFDM \cite{9123984}. Furthermore, an improved least-squares detection scheme with QR factorization (mLSQR) \cite{10261239} and an alternating optimization method for multidimensional resource allocation (AO-MRA) \cite{11164940} have been proposed for NOMA-OTFS. Iterative SCMA detection for AFDM \cite{10566604} and GNN-based detection for NOMA-AFDM \cite{11557489} have also been investigated.
	
	\textcolor{black}{However, conventional power-domain NOMA, originally developed for single-antenna settings, may suffer from multiplexing gain loss when directly extended to multi-antenna systems, due to the inefficient utilization of spatial dimensions and the requirement for users to fully decode and cancel the interference from other users \cite{9451194,10720669}.}  Power-domain NOMA mainly improves spectral efficiency through power allocation and successive interference cancellation (SIC), but its spatial interference-management capability remains limited \cite{9831440}. Similarly, sparse code multiple access (SCMA) and NOMA rely on superimposing multi-user signals over shared resources, which differs from multi-antenna transmission that separates streams and suppresses interference in the spatial domain \cite{10471302,10032203}.
	
	Rate-splitting multiple access (RSMA) \cite{mao2018rate,7470942} provides a more flexible non-orthogonal transmission strategy \cite{10858168,10855342,10299599,10312769}. In downlink RSMA, each user’s message is divided into common and private parts, with all common parts jointly encoded into a common stream. Each receiver first decodes and removes the common stream before detecting its private stream. By converting part of the inter-user interference into decodable information, RSMA actively manages interference while exploiting the spatial multiplexing capability of multi-antenna systems \cite{9451194,10032203}. Compared with space-division multiple access (SDMA), RSMA also provides greater robustness by partially decoding interference \cite{9831440}.
	
	Cooperative communication further complements RSMA \cite{10038476}. After decoding the common stream, user terminals can forward cooperative information to cell-edge users, thereby improving common-stream reliability, mitigating severe path loss, and extending coverage \cite{10038476,9831440}. Cooperative integrated sensing and communication (ISAC)-RSMA systems have been investigated in \cite{11106465,10032141,11153084}, while joint precoding and rate-allocation optimization has been applied to improve fairness in dense visible-light communication systems \cite{11072172}.

\textcolor{black}{In this paper, we investigate cooperative RSMA-AFDM transmission with user-dependent chirp design. A PEP analysis is first developed to characterize the impact of chirp parameters on cooperative common-stream detection, based on which a PEP-guided cooperative effective-distance optimization scheme is proposed to improve the channel-averaged detection reliability. Furthermore, two distributed cooperative EP-based detection schemes, namely decision fusion and belief consensus, are developed to further exploit the benefits of multiuser cooperation.}
	
	The main contributions of this paper are summarized below:
	\begin{itemize}
		\item{\textcolor{black}{We propose a cooperative RSMA-AFDM system. Cooperative RSMA effectively addresses the multiple access problem in AFDM while fully exploiting the multiplexing gain provided by the spatial degrees of freedom in multi-antenna systems to enhance spectral efficiency. Furthermore, the AFDM chirp parameters provide a channel-aware design degree of freedom for optimizing the cooperative effective distance.}}
		\textcolor{black}{\item{We derive a conditional PEP expression for cooperative
			common-stream detection, explicitly retaining the common-stream
			precoder in the effective-distance term. For MRT precoding, the
			expression is further expanded to identify the individual-user and
			cross-user contributions to the cooperative effective distance.}
			\item{\textcolor{black}{We establish a PEP-guided user-dependent chirp parameter design for
			multiuser cooperative RSMA-AFDM. By anchoring one user's chirp
			parameter to the conventional AFDM choice and jointly selecting the
			remaining parameters over their AFDM-admissible sets, the proposed
			design directly optimizes the channel-averaged PEP through the
			cooperative effective-distance criterion.}}}
		\item{\textcolor{black}{We propose two distributed cooperative algorithms for common stream detection. The first decision fusion--based method combines the information from local EP detection with information broadcast by other users via effective-SNR-based belief fusion to achieve a globally consistent distributed cooperative estimate. The second belief consensus-based method exchanges and sums the external Gaussian likelihood parameters across users while retaining a single shared EP site across iterations. Through iterative exchanges, distributed nodes collaboratively refine their beliefs and converge to a consistent and improved detection decision.}}
	\end{itemize}

	The remainder of this paper is organized as follows.
	Section II presents the cooperative RSMA-AFDM system model.
	\textcolor{black}{Section III analyzes the cooperative effective distance and establishes the PEP objective for chirp-parameter optimization.
	Section IV derives the user-specific $c_1$.}
	Section V introduces the decision fusion-based and belief consensus-based distributed cooperative detectors. \textcolor{black}{Section VI presents the simulation results and complexity analysis}, and Section VII concludes the paper.

	\section{Cooperative RSMA-AFDM System Model}
\begin{figure*}
		\centering
		\includegraphics[width=0.8\textwidth]{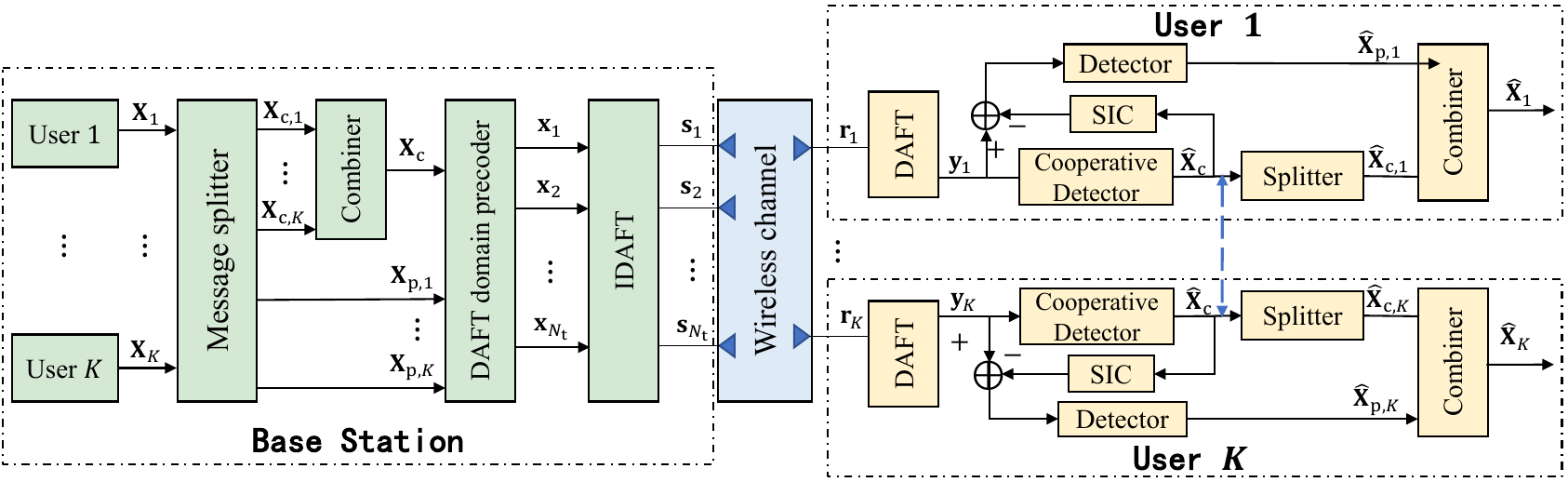}
		\caption{Block diagram of the downlink cooperative RSMA-AFDM system.}
		\label{str}
	\end{figure*}

	In this section, we present the system model for the downlink cooperative RSMA-AFDM transmission.
	As shown in Fig. \ref{str}, at the base station, the transmitted symbols  are first split into common streams and private streams. The common and private streams are then precoded in the DAFT domain and converted into time-domain signals using AFDM. At the receiver, the time-domain received signals are converted to the DAFT domain through AFDM demodulation.   

	Let $N$ represent the number of symbols per frame. The transmitted symbol for User $k \in \{1, 2,..., K\}$ is expressed as $\mathbf{X}_k$, where $K$ is the number of users. First, $\mathbf{X}_k$ is divided into a common part $\mathbf{X}_{\mathrm{c},k}$ and a private part $\mathbf{X}_{\mathrm{p},k}$. The common parts of all users are concatenated into $\mathbf{X}_\mathrm{c} = [\mathbf{X}^{T}_{\mathrm{c},1},\mathbf{X}^{T}_{\mathrm{c},2},..., \mathbf{X}^{T}_{\mathrm{c},K}]^{T}$ and precoded in the DAFT domain to obtain the common stream, where $(\cdot)^{T}$ is the matrix transpose. For simplicity in subsequent analysis, we set $\mathbf{X}_\mathrm{c}\in \mathbb{C}^{N \times 1}$ and $\mathbf{X}_{\mathrm{p},k}\in \mathbb{C}^{N \times 1}$, where $\mathbb{C}$ denotes the set of complex numbers. If the coded and modulated sequence does not exactly contain $N$ symbols, excessive symbols are buffered for the next frame, while zero padding is applied when the sequence is shorter than $N$. The private part of each user is individually precoded in the DAFT domain. The final precoded DAFT domain transmitted symbols are expressed as
	\begin{equation}
		\mathbf{x}_{\mathrm{vec}}=\sqrt{\beta_\mathrm{c}}\mathbf{w}_\mathrm{c}\mathbf{X}_\mathrm{c}+\sum_{k=1}^K\sqrt{\beta_{\mathrm{p},k}}\mathbf{w}_{\mathrm{p},k}\mathbf{X}_{\mathrm{p},k},
		\label{x}
	\end{equation}
	where $\mathbf{x}_{\mathrm{vec}} = [\mathbf{x}^{T}_{1},\mathbf{x}^{T}_{2},...,\mathbf{x}^{T}_{N_t}]^{T}\in \mathbb{C}^{N_tN \times 1}$; $N_t$ is the number of antennas; ${\beta_\mathrm{c}}$ and $\beta_{\mathrm{p},k}$ denote the power allocated to the common stream, and the private stream of User $k$, respectively, with ${\beta_\mathrm{c}}+\sum_{k=1}^K{\beta_{\mathrm{p},k}} = 1$; $\mathbf{w}_\mathrm{c} \in \mathbb{C}^{N_tN \times N}$ is the precoding matrix for the common stream; $\mathbf{w}_{\mathrm{p},k}$  represents the precoding matrix for the private stream of User $k$.  Without loss of generality, we employ maximum ratio transmission (MRT) precoding for the common stream and zero forcing (ZF) precoding for the private streams \cite{9967957}. The MRT precoding for the common stream is expressed as
	\begin{equation}
		\mathbf{w}_\mathrm{c}=\frac{
			\sum_{k=1}^K\hat{\mathbf{H}}_{k}^{H}}{\left\|\sum_{k=1}^K\mathbf{\hat{\mathbf{H}}}_{k}^{H}\right\|},
		\label{wc}
	\end{equation}
	where $\hat{\mathbf{H}}_{k}\in \mathbb{C}^{N \times N_tN}$ denotes the estimated DAFT domain equivalent channel of User $k$; $(\cdot)^{\mathrm{H}}$ is the Hermitian operation; $\left\|\cdot\right\|$ is the Euclidean norm. The ZF precoding for the private streams needs to calculate the pseudo-inverse of matrix $\hat{\mathbf{H}}=[\hat{\mathbf{H}}^{T}_{1},\hat{\mathbf{H}}^{T}_{2},...,\hat{\mathbf{H}}^{T}_{K}]^{T}\in \mathbb{C}^{NK \times N_tN}$, which can be expressed as 
	\begin{equation}
		\textcolor{black}{\mathbf{D}=\frac{\hat{\mathbf{H}}^{H}\left(\hat{\mathbf{H}}\hat{\mathbf{H}}^{H}\right)^{-1}}{\left\|\hat{\mathbf{H}}^{H}\left(\hat{\mathbf{H}}\hat{\mathbf{H}}^{H}\right)^{-1}\right\|_{\mathrm F}}.}
	\end{equation}
	\textcolor{black}{This inverse-based ZF expression assumes $N_t\geq K$ and $\operatorname{rank}(\hat{\mathbf H})=KN$.}
	User $k$'s private stream precoding matrix $\mathbf{w}_{\mathrm{p},k}$ is written as
	\begin{equation}
		\mathbf{w}_{\mathrm{p},k}=\mathbf{D}[:,(k-1)N+1:kN].
	\end{equation}
	
	The DAFT-domain vector $\mathbf{x}_{\mathrm{vec}}$ is reshaped as
	$\mathbf{x}
	=
	[\mathbf{x}_1,\mathbf{x}_2,\ldots,\mathbf{x}_{N_t}]
	\in\mathbb{C}^{N\times N_t}.$
	Since each transmit antenna emits a single composite AFDM waveform in one frame, the common and private components superimposed on the same antenna share one transmit-side chirp-parameter pair $(c_{1,\mathrm t},c_{2,\mathrm t})$. The same transmit-side IDAFT matrix is applied columnwise to all transmit antennas, yielding
	\begin{equation}
	\mathbf{s}
	=
	\mathbf{A}_{\mathrm t}^{H}\mathbf{x},
	\label{eq:idaft_tx}
\end{equation}
	where $\mathbf{A}_{\mathrm t}^{H}
	=
	\boldsymbol{\Lambda}_{c_{1,\mathrm t}}^{H}
	\mathbf{F}^{H}
	\boldsymbol{\Lambda}_{c_{2,\mathrm t}}^{H}
	\in\mathbb{C}^{N\times N}$
	is the transmit-side IDAFT matrix; $\mathbf{F}\in\mathbb{C}^{N\times N}$ is the normalized DFT matrix with entries $e^{-j2\pi mn/N}/\sqrt{N}$, and
	\begin{equation}
	\boldsymbol{\Lambda}_{c_{1,\mathrm t}}
	=
	\operatorname{diag}
	\left(
	e^{-j2\pi c_{1,\mathrm t}n^2},
	n=0,\ldots,N-1
	\right),
\end{equation}
	\begin{equation}
	\boldsymbol{\Lambda}_{c_{2,\mathrm t}}
	=
	\operatorname{diag}
	\left(
	e^{-j2\pi c_{2,\mathrm t}n^2},
	n=0,\ldots,N-1
	\right).
\end{equation}
	The parameters $c_{1,\mathrm t}$ and $c_{2,\mathrm t}$ are the transmit-side chirp parameters, and $\mathbf{s}=[\mathbf{s}_{1},\mathbf{s}_{2},\ldots,\mathbf{s}_{N_t}]\in\mathbb{C}^{N\times N_t}$ is the time-domain transmit signal matrix. After inserting the chirp-periodic prefix (CPP), $\mathbf{s}$ is transmitted to all users.
	
	At the receiver, the time-domain received signal of User $k$ can be expressed as
	\begin{equation}
		{r_{k}}[n]=\sum_{i=1}^{N_t}\sum_{l=0}^\infty{s}_i[n-l]\sum_{p=1}^Ph_{p}^{[k,i]}e^{-j2\pi f_{p}^{[k,i]}n}\delta(l-l_{p}^{[k,i]})+{z_k}[n],
		\label{tre}
	\end{equation}
	where $\delta(\cdot)$ is the Dirac delta function; $h_{p}^{[k,i]}$, $f_{p}^{[k,i]}$, and $l_{p}^{[k,i]}$ represent the gain, Doppler shift index, and delay index of the $p$-th path from the $i$-th antenna to User $k$'s channel, respectively; $P$ denotes the number of paths; ${z_k}[n]$ denotes the zero-mean complex additive white Gaussian noise (AWGN) with variance $\sigma^2_k$. The  matrix form of \eqref{tre} can be expressed as
	\begin{equation}
	\mathbf{r}_k=\sum_{i=1}^{N_t}\mathbf{H}_\mathrm{T}^{[k,i]}\mathbf{s}_i+\mathbf{z}_k,
\end{equation}
	where $\mathbf{H}_\mathrm{T}^{[k,i]}$ denotes the time-domain equivalent channel from the $i$-th antenna to the $k$-th user.
	
	User $k$ performs AFDM demodulation using its own chirp parameters $(c_{1,k},c_{2,k})$. Therefore, its DAFT-domain received vector is
	\begin{equation}
	\mathbf{y}_k
	=
	\mathbf{A}_k\mathbf{r}_k
	=
	\boldsymbol{\Lambda}_{c_{2,k}}
	\mathbf{F}
	\boldsymbol{\Lambda}_{c_{1,k}}
	\mathbf{r}_k.
	\label{eq:daft_rx_k}
\end{equation}

	By stacking the received signals from the $K$ users, we obtain
	\begin{equation}\mathbf{y}=\mathbf{H}\mathbf{x}_{\mathrm{vec}}+\mathbf{z},\label{daftio}
	\end{equation}
	where $\mathbf{y} = [\mathbf{y}^{T}_{1},\mathbf{y}^{T}_{2},..., \mathbf{y}^{T}_{K}]^{T}\in \mathbb{C}^{NK \times 1}$;  $\mathbf{z} = [\mathbf{z}^{T}_{1},\mathbf{z}^{T}_{2},...,\mathbf{z}^{T}_{K}]^{T}\in \mathbb{C}^{NK \times 1}$; $\mathbf{H}\in \mathbb{C}^{NK \times N_tN}$ is the DAFT domain equivalent channel matrix and can be expressed as
	\begin{equation}\mathbf{H}=
		\begin{bmatrix}
			\mathbf{H}^{[1,1]} & \mathbf{H}^{[1,2]} & \cdots & \mathbf{H}^{[1,N_t]} \\
			\mathbf{H}^{[2,1]} & \mathbf{H}^{[2,2]} & \cdots & \mathbf{H}^{[2,N_t]} \\
			\vdots & \vdots & \ddots & \vdots \\
			\mathbf{H}^{[K,1]} & \mathbf{H}^{[K,2]} & \cdots & \mathbf{H}^{[K,N_t]} \\
		\end{bmatrix}\in\mathbb{C}^{NK\times N_tN},\end{equation}
	where
	\begin{equation}
		\mathbf{H}^{[k,i]}
		=
		\sum_{p=1}^{P}
		h_{p}^{[k,i]}
		\mathbf{A}_k
		\mathbf{\Gamma}_{\mathrm{CPP},p}^{[k,i]}
		\mathbf{\Delta}_{f_{p}^{[k,i]}}
		\mathbf{\Pi}^{l_{p}^{[k,i]}}
		\mathbf{A}_{\mathrm t}^{H},
		\label{hdaft}
	\end{equation}
	denotes the DAFT domain equivalent channel from the $i$-th antenna to User $k$, where $\mathbf{\Delta}_{f_{p}^{[k,i]}}\triangleq\mathrm{diag}(e^{-j{2\pi} f_{p}^{[k,i]}n},n=0,1,...,N-1)\in\mathbb{C}^{N\times N}$ is the spreading matrix induced by the Doppler shifts, 
	\begin{equation}
		\mathbf{\Gamma}_{\mathrm{CPP},p}^{[k,i]}
		=
		\operatorname{diag}
		\left(
		\begin{cases}
			e^{-j2\pi c_{1,\mathrm t}
				\left(N^2-2N(l_{p}^{[k,i]}-n)\right)},
			& n<l_{p}^{[k,i]},\\
			1, & n\geq l_{p}^{[k,i]}
		\end{cases}
		\right)
		\label{cpp}
	\end{equation}
	is the equivalent CPP matrix \cite{10711268}; $\mathbf{\Pi}\in\mathbb{C}^{N\times N}$ represents the forward cyclic shift matrix \cite{13}, given by
	\begin{equation}
		\mathbf{\Pi}=\begin{bmatrix}0&\cdots&0&1\\1&\cdots&0&0\\&\ddots&\ddots&\\0&\cdots&1&0\end{bmatrix}_{N\times N}.
	\end{equation}
	When $(c_{1,k},c_{2,k})\neq(c_{1,\mathrm t},c_{2,\mathrm t})$, the transmit and receive AFDM bases are related by the deterministic cross-basis matrix
	$\mathbf{C}_{k\leftarrow\mathrm t}
	=
	\mathbf{A}_k\mathbf{A}_{\mathrm t}^{H}.$
	This known term is already included in $\mathbf{H}^{[k,i]}$ and can be accounted for during common-stream detection, resulting in only a minor residual performance loss.
	
	Substituting (\ref{x}) into (\ref{daftio}), we have
	\begin{equation}
		\begin{aligned}
			\mathbf{y}&=\mathbf{H}(\sqrt{\beta_\mathrm{c}}\mathbf{w}_\mathrm{c}\mathbf{X}_\mathrm{c}+\sum_{k=1}^{K}\sqrt{\beta_{\mathrm{p},k}}\mathbf{w}_{\mathrm{p},k}\mathbf{X}_{\mathrm{p},k})+\mathbf{z} \\& = \sqrt{\beta_\mathrm{c}}\mathbf{G}_\mathrm{c}\mathbf{X}_\mathrm{c}+\sum_{k=1}^{K}\sqrt{\beta_{\mathrm{p},k}}\textcolor{black}{\mathbf{H}\mathbf{w}_{\mathrm{p},k}}\mathbf{X}_{\mathrm{p},k}+\mathbf{z},
		\end{aligned}
		\label{daftrsmaio}
	\end{equation}
	\textcolor{black}{where $\mathbf{G}_\mathrm{c} = \mathbf{H}\mathbf{w}_\mathrm{c}\in\mathbb{C}^{NK\times N}$ is the equivalent channel matrix of the common stream.}
	
	Next, the transmitted symbols need to be detected. The detection process consists of two steps. First, all user nodes cooperatively detect the common stream. Subsequently, each user eliminates the common stream components and independently detects the respective private streams. 
	Since all users receive the same common stream, the cooperative RSMA-AFDM system can integrate broadcast information from other users to enhance the detection reliability through information fusion. Subsequently, using SIC scheme, the decoded common stream component is removed from the received signal. Finally, the private streams are detected. Given that ZF precoding is applied to the private streams, the receiver can recover the private streams using a simple linear equalization scheme. Ultimately, by combining the user-specific common stream portion with the private stream, the detection is performed.

	\color{black}
	\section{PEP Analysis}
	\label{sec:pep_analysis}
	
	This section derives the PEP of the
	cooperative common stream and identifies how the user-dependent AFDM
	chirp parameters enter the PEP. The analysis focuses on the two-user
	case, while the formulation can be extended to more users by stacking
	their observations. The
	following derivation explicitly retains the common-stream precoder in
	the effective distance. This is important because the receive-side
	chirp parameters affect the PEP not only through the DAFT-domain
	representation, but also through the precoder constructed from the
	DAFT-domain equivalent channels.
	
	\subsection{Conditional PEP Derivation}
	
	For the common-stream detection stage, the received signal of User $k$
	is expressed as
	\begin{equation}
		\mathbf y_k
		=
		\sqrt{\beta_{\mathrm c}}\,
		\mathbf G_{\mathrm c,k}\mathbf X_{\mathrm c}
		+\boldsymbol{\psi}_k,
		\label{eq:pep_user_signal}
	\end{equation}
	where
	\begin{equation}
		\mathbf G_{\mathrm c,k}
		=
		\mathbf H_k\mathbf w_{\mathrm c}
		\in\mathbb C^{N\times N}
	\end{equation}
	is the effective common-stream channel, and
	$\boldsymbol{\psi}_k$ contains the private-stream interference and
	receiver noise. The stacked cooperative observation is
	\begin{equation}
		\mathbf y_{\mathrm{co}}
		=
		\sqrt{\beta_{\mathrm c}}\,
		\mathbf G_{\mathrm{co}}\mathbf X_{\mathrm c}
		+\boldsymbol{\psi}_{\mathrm{co}},
		\label{eq:pep_stacked_signal}
	\end{equation}
	with
	\begin{align}
		\mathbf G_{\mathrm{co}}
		=
		\begin{bmatrix}
			\mathbf G_{\mathrm c,1}\\
			\mathbf G_{\mathrm c,2}
		\end{bmatrix}
		=
		\begin{bmatrix}
			\mathbf H_1\\
			\mathbf H_2
		\end{bmatrix}
		\mathbf w_{\mathrm c},   \quad                                   
		\boldsymbol{\psi}_{\mathrm{co}}
		=
		\begin{bmatrix}
			\boldsymbol{\psi}_1\\
			\boldsymbol{\psi}_2
		\end{bmatrix}.
	\end{align}
	Here, $\mathbf H_k$ is the DAFT-domain equivalent channel defined in
	Section II and therefore depends on the receive-side chirp parameters.
	
	Consider two distinct common-stream codewords
	$\mathbf X_{\mathrm c}$ and $\widehat{\mathbf X}_{\mathrm c}$. Define
	the error vector
	\begin{equation}
		\boldsymbol{\delta}
		=
		\mathbf X_{\mathrm c}-\widehat{\mathbf X}_{\mathrm c}
		\neq\mathbf 0.
		\label{eq:pep_delta}
	\end{equation}
	The difference between the corresponding noiseless cooperative
	received vectors is
	\begin{equation}
		\Delta\boldsymbol{\mu}_{\mathrm{co}}
		=
		\sqrt{\beta_{\mathrm c}}\,
		\mathbf G_{\mathrm{co}}\boldsymbol{\delta}.
		\label{eq:pep_mean_difference}
	\end{equation}
	
	Assume that the aggregate disturbance is modeled as a circularly
	symmetric complex Gaussian vector,
	\begin{equation}
		\boldsymbol{\psi}_{\mathrm{co}}
		\sim\mathcal{CN}(\mathbf 0,\mathbf R_{\psi}),
		\qquad \mathbf R_{\psi}\succ\mathbf 0,
		\label{eq:pep_disturbance}
	\end{equation}
	where
	\begin{equation}
		\mathbf R_{\psi}
		=
		\mathbb E\left[
		\boldsymbol{\psi}_{\mathrm{co}}
		\boldsymbol{\psi}_{\mathrm{co}}^H
		\right].
	\end{equation}
	For coherent maximum-likelihood detection with the correct covariance
	matrix, the conditional PEP is
	\begin{equation}
		{
			P\!\left(
			\mathbf X_{\mathrm c}\rightarrow
			\widehat{\mathbf X}_{\mathrm c}
			\mid\mathbf H_1,\mathbf H_2
			\right)
			=
			Q\!\left(
			\sqrt{
				\frac{\beta_{\mathrm c}}{2}
				\boldsymbol{\delta}^H
				\mathbf G_{\mathrm{co}}^H
				\mathbf R_{\psi}^{-1}
				\mathbf G_{\mathrm{co}}
				\boldsymbol{\delta}}
			\right).
		}
		\label{eq:pep_general}
	\end{equation}
	The corresponding whitened effective distance is
	\begin{equation}
		d_{\mathrm{wh}}^2(\boldsymbol{\delta})
		=
		\boldsymbol{\delta}^H
		\mathbf G_{\mathrm{co}}^H
		\mathbf R_{\psi}^{-1}
		\mathbf G_{\mathrm{co}}
		\boldsymbol{\delta},
		\label{eq:dwh}
	\end{equation}
	so that
	\begin{equation}
		P_{\mathrm{PEP}}(\boldsymbol{\delta})
		=
		Q\!\left(
		\sqrt{\frac{\beta_{\mathrm c}}{2}
			d_{\mathrm{wh}}^2(\boldsymbol{\delta})}
		\right).
		\label{eq:pep_dwh}
	\end{equation}
	If the aggregate disturbance is white with
	$\mathbf R_{\psi}=\sigma^2\mathbf I$, this reduces to
	\begin{equation}
		{
			P_{\mathrm{PEP}}(\boldsymbol{\delta})
			=
			Q\!\left(
			\sqrt{\frac{\beta_{\mathrm c}}{2\sigma^2}
				\left\|\mathbf G_{\mathrm{co}}\boldsymbol{\delta}\right\|^2}
			\right).
		}
		\label{eq:pep_white}
	\end{equation}
	
		The conditional PEP in \eqref{eq:pep_general} can be further related
		to the path-gain statistics. \textcolor{black}{To characterize the path-dependent effective-distance exponent,} first
	consider the structural PEP analysis in which the common-stream
	precoder is fixed over the considered pairwise error event. For the
	$k$th user and the $i$th transmit antenna, let
	$\mathbf G_{\mathrm c,p}^{[k,i]}$ denote the effective common-stream
	matrix associated with the $p$th propagation path, including the
	corresponding path-independent channel operators and the fixed
	precoder. The received difference associated with this link can be
	written as
	\begin{equation}
		\sqrt{\beta_{\mathrm c}}
		\sum_{p=1}^{P}h_p^{[k,i]}
		\mathbf G_{\mathrm c,p}^{[k,i]}\boldsymbol{\delta}
		=
		\sqrt{\beta_{\mathrm c}}\,
		\boldsymbol{\Phi}^{[k,i]}(\boldsymbol{\delta})
		\mathbf h^{[k,i]},
		\label{eq:pep_path_decomposition}
	\end{equation}
	where
	\begin{align}
		\mathbf h^{[k,i]}
		&=
		[h_1^{[k,i]},\ldots,h_P^{[k,i]}]^T,\\
		\boldsymbol{\Phi}^{[k,i]}(\boldsymbol{\delta})
		&=
		[\mathbf G_{\mathrm c,1}^{[k,i]}\boldsymbol{\delta},
		\ldots,
		\mathbf G_{\mathrm c,P}^{[k,i]}\boldsymbol{\delta}].
		\label{eq:Phi_link}
	\end{align}
	Collecting the links of each user gives
	\begin{align}
		\mathbf h_k
		&=
		[(\mathbf h^{[k,1]})^T,\ldots,
		(\mathbf h^{[k,N_t]})^T]^T,\\
		\boldsymbol{\Phi}_k(\boldsymbol{\delta})
		&=
		[\boldsymbol{\Phi}^{[k,1]}(\boldsymbol{\delta}),
		\ldots,
		\boldsymbol{\Phi}^{[k,N_t]}(\boldsymbol{\delta})].
	\end{align}
	For the two-user cooperative observation, define
	\begin{equation}
		\mathbf h_{\mathrm{co}}
		=
		\begin{bmatrix}\mathbf h_1\\\mathbf h_2\end{bmatrix},
		\qquad
		\boldsymbol{\Phi}_{\mathrm{co}}(\boldsymbol{\delta})
		=
		\operatorname{blkdiag}
		\left(
		\boldsymbol{\Phi}_1(\boldsymbol{\delta}),
		\boldsymbol{\Phi}_2(\boldsymbol{\delta})
		\right).
		\label{eq:Phi_co_block}
	\end{equation}
	The whitened pairwise observation difference is then
	\begin{equation}
		\check{\Delta\mathbf y}_{\mathrm{co}}
		=
		\sqrt{\beta_{\mathrm c}}\,
		\mathbf R_{\psi}^{-1/2}
		\boldsymbol{\Phi}_{\mathrm{co}}(\boldsymbol{\delta})
		\mathbf h_{\mathrm{co}}
		+\check{\boldsymbol{\psi}}_{\mathrm{co}}.
		\label{eq:pep_whitened_path}
	\end{equation}
	
	Assume independent zero-mean complex Gaussian path gains with a positive diagonal covariance matrix
	\begin{equation}
		\mathbf h_{\mathrm{co}}\sim
		\mathcal{CN}(\mathbf 0,\mathbf R_h),
		\label{eq:path_gain_distribution}
	\end{equation}
	where $\mathbf R_h$ is positive definite on the active path-gain
	dimensions. For a fixed precoder and a fixed error vector, define
	\begin{equation}
		\mathbf M_{\delta}
		=
		\mathbf R_h^{1/2}
		\boldsymbol{\Phi}_{\mathrm{co}}^H(\boldsymbol{\delta})
		\mathbf R_{\psi}^{-1}
		\boldsymbol{\Phi}_{\mathrm{co}}(\boldsymbol{\delta})
		\mathbf R_h^{1/2}.
		\label{eq:M_delta}
	\end{equation}
	Using Craig's representation of the Gaussian $Q$-function and the
	moment-generating function of a quadratic form in a complex Gaussian
	vector, the PEP averaged over the path gains is
	\begin{equation}
		{
			\overline P_{\mathrm{PEP}}(\boldsymbol{\delta})
			=
			\frac{1}{\pi}
			\int_{0}^{\pi/2}
			\det\!\left(
			\mathbf I+
			\frac{\beta_{\mathrm c}}{4\sin^2\theta}
			\mathbf M_{\delta}
			\right)^{-1}
			d\theta .
		}
		\label{eq:pep_determinant}
	\end{equation}
	Let $r_{\delta}=\operatorname{rank}(\mathbf M_{\delta})$. At high SNR,
	provided that the nonzero eigenvalues of $\mathbf M_{\delta}$ remain
	strictly positive and the precoder is fixed with respect to the
	averaged path-gain vector, the dominant PEP term decays with SNR as
	\begin{equation}
		\overline P_{\mathrm{PEP}}(\boldsymbol{\delta})
		\doteq \rho^{-r_{\delta}},
		\label{eq:pep_rank_exponent}
	\end{equation}
		where $\doteq$ denotes equality in the SNR-exponent sense. Thus, the
	\textcolor{black}{rank of the whitened path-response matrix determines the
		high-SNR PEP decay exponent} for the fixed-precoder structural model.
	
	In the actual implementation,
	$\mathbf w_{\mathrm c}$ is recomputed from the estimated channel
	matrices for each channel realization. Therefore, the path-response
	matrix can itself depend on the path gains, and the determinant
	averaging in \eqref{eq:pep_determinant} is not generally an exact
	fading average for that adaptive precoder. In the numerical
	optimization, the conditional PEP is instead evaluated using the
	precoder recomputed for each channel realization and then averaged
	over the channel samples. 
	
	\subsection{Explicit Dependence on the Chirp Parameters}
	
	The DAFT-domain equivalent channel of User $k$ is
	\begin{equation}
		\mathbf H_k
		=
		\mathbf A_k\mathbf H_{\mathrm T,k}\mathbf A_{\mathrm t}^H,
		\label{eq:pep_Hk}
	\end{equation}
	where the receive-side DAFT matrix is
	\begin{equation}
		\mathbf A_k
		=
		\boldsymbol{\Lambda}_{c_{2,k}}
		\mathbf F
		\boldsymbol{\Lambda}_{c_{1,k}}.
		\label{eq:pep_Ak}
	\end{equation}
	The common-stream precoder is constructed from the estimated
	DAFT-domain channels. In general, it can be written as
	\begin{equation}
		\mathbf w_{\mathrm c}
		=
		\mathcal P(\mathbf H_1,\mathbf H_2),
		\label{eq:pep_precoder_general}
	\end{equation}
	where $\mathcal P(\cdot)$ denotes the adopted precoding rule, including
	the normalization required by the transmit-power constraint. Therefore,
	\begin{equation}
		\mathbf G_{\mathrm{co}}
		=
		\begin{bmatrix}\mathbf H_1\\\mathbf H_2\end{bmatrix}
		\mathcal P(\mathbf H_1,\mathbf H_2)
		\label{eq:pep_Gchirp}
	\end{equation}
	is a function of $(c_{1,1},c_{1,2})$.
	
	For the MRT common-stream precoder adopted in the analytical model,
	\begin{equation}
		\mathbf w_{\mathrm c}
		=
		\frac{\mathbf H_1^H+\mathbf H_2^H}
		{\|\mathbf H_1^H+\mathbf H_2^H\|_F},
		\label{eq:pep_mrt}
	\end{equation}
	where a scalar factor can be added to satisfy the prescribed common
	stream power. Define
	\begin{equation}
		\mathbf B_{\mathrm{co}}
		=
		\begin{bmatrix}\mathbf B_1\\\mathbf B_2\end{bmatrix},
		\qquad
		\mathbf S
		=
		\mathbf B_1^H\mathbf A_1^H+
		\mathbf B_2^H\mathbf A_2^H,
		\label{eq:pep_S}
	\end{equation}
	where 
	$\mathbf B_k=\mathbf H_{\mathrm T,k}\mathbf A_{\mathrm t}^{H}$.
	Then $\mathbf w_{\mathrm c}=\mathbf S/\|\mathbf S\|_F$, and under
	white disturbance the effective distance in \eqref{eq:dwh} becomes
	\begin{equation}
		d_{\mathrm{eff}}^2(\boldsymbol{\delta})
		=
		\frac{
			\boldsymbol{\delta}^H
			\mathbf S^H\mathbf B_{\mathrm{co}}^H
			\mathbf B_{\mathrm{co}}\mathbf S
			\boldsymbol{\delta}}
		{\|\mathbf S\|_F^2}.
		\label{eq:pep_mrt_distance}
	\end{equation}
	Consequently,
	\begin{equation}
		{
			P_{\mathrm{PEP}}(\boldsymbol{\delta})
			=
			Q\!\left(
			\sqrt{\frac{\beta_{\mathrm c}}{2\sigma^2}
				\frac{
					\boldsymbol{\delta}^H
					\mathbf S^H\mathbf B_{\mathrm{co}}^H
					\mathbf B_{\mathrm{co}}\mathbf S
					\boldsymbol{\delta}}
				{\|\mathbf S\|_F^2}}
			\right).
		}
		\label{eq:pep_mrt_final}
	\end{equation}
	Equation \eqref{eq:pep_mrt_final} explicitly shows how the receive-side
	chirp parameters affect the PEP through the channel-dependent MRT
	combination $\mathbf S$ and the resulting effective transmit direction.
	
	For completeness, write
	\begin{equation}
		\mathbf U_1=\mathbf B_1^H\mathbf A_1^H,\qquad
		\mathbf U_2=\mathbf B_2^H\mathbf A_2^H,
		\qquad \mathbf S=\mathbf U_1+\mathbf U_2.
	\end{equation}
	With
	$\mathbf Q_{\mathrm{co}}=\mathbf B_{\mathrm{co}}^H\mathbf B_{\mathrm{co}}$,
	the numerator matrix of \eqref{eq:pep_mrt_distance} can be expanded as
	\begin{align}
		\mathbf S^H\mathbf Q_{\mathrm{co}}\mathbf S
		={}&
		\mathbf U_1^H\mathbf Q_{\mathrm{co}}\mathbf U_1
		+\mathbf U_2^H\mathbf Q_{\mathrm{co}}\mathbf U_2
		\nonumber\\
		&+
		\mathbf U_1^H\mathbf Q_{\mathrm{co}}\mathbf U_2
		+\mathbf U_2^H\mathbf Q_{\mathrm{co}}\mathbf U_1.
		\label{eq:pep_cross_expansion}
	\end{align}
	The last two terms describe the cross-user contribution to the
	cooperative effective distance. Their effect can be constructive or
	destructive, depending on the channel realization and chirp parameters.
	Therefore, the PEP objective captures more information than a
	path-location overlap count alone.
	
	\subsection{Average PEP Criterion}
	
	Let $\mathcal X=\mathcal A^N$ denote the common-stream codebook, where  $\mathcal{A}$ is the modulation constellation set. For
	equiprobable transmitted codewords, define the average pairwise-error
	sum as
	\begin{align}
		\overline P_{\mathrm{PEP}}
		(\mathbf c_1)
		={}&
		\mathbb E_{\mathbf H_1,\mathbf H_2}
		\Bigg[
		\frac{1}{|\mathcal X|}
		\sum_{\mathbf X_{\mathrm c}\in\mathcal X}
		\sum_{\substack{\widehat{\mathbf X}_{\mathrm c}\in\mathcal X\\
				\widehat{\mathbf X}_{\mathrm c}\ne\mathbf X_{\mathrm c}}}\times
		\nonumber\\[-1mm]
		&\qquad
		P_{\mathrm{PEP}}\!
		\left(
		\mathbf X_{\mathrm c}\rightarrow
		\widehat{\mathbf X}_{\mathrm c}
		\mid\mathbf H_1,\mathbf H_2
		\right)
		\Bigg].
		\label{eq:average_pep}
	\end{align}
	This averaged pairwise-error sum is a PEP-based design metric and can
	be used in a union-bound analysis. 
	
	\subsection{Extension to More Than Two Users}
	\label{subsec:pep_multiuser_extension}
	
	The above PEP formulation can be directly extended to a cooperative
	system with $K\geq 3$ users. In this case, the received observations
	from all participating users are stacked to form
	\begin{equation}
		\mathbf y_{\mathrm{co}}
		=
		\sqrt{\beta_{\mathrm c}}\,
		\mathbf G_{\mathrm{co}}^{(K)}
		\mathbf X_{\mathrm c}
		+\boldsymbol{\psi}_{\mathrm{co}},
		\label{eq:pep_multiuser_signal}
	\end{equation}
	where
\begin{align}
	\mathbf G_{\mathrm{co}}^{(K)}
	&=
	\left[\mathbf H_1^T,\ldots,\mathbf H_K^T\right]^T
	\mathbf w_{\mathrm c}
	=
	\left[\mathbf G_{\mathrm c,1}^T,\ldots,
	\mathbf G_{\mathrm c,K}^T\right]^T,
	\label{eq:pep_multiuser_G}\\
	\mathbf R_{\psi}^{(K)}
	&=
	\mathbb E\left[
	\boldsymbol{\psi}_{\mathrm{co}}
	\boldsymbol{\psi}_{\mathrm{co}}^H
	\right].
\end{align}
	Here, the common-stream precoder is jointly designed from all users'
	DAFT-domain equivalent channels, i.e.,
	\begin{equation}
		\mathbf w_{\mathrm c}
		=
		\mathcal P(\mathbf H_1,\mathbf H_2,\ldots,\mathbf H_K),
		\qquad
		\mathbf H_k=\mathbf A_k\mathbf B_k.
		\label{eq:pep_multiuser_precoder}
	\end{equation}
	For a pairwise error event
	$\boldsymbol\delta=\mathbf X_{\mathrm c}
	-\widehat{\mathbf X}_{\mathrm c}$, the conditional PEP is therefore
	\begin{equation}
		{
			P_{\mathrm{PEP}}^{(K)}(\boldsymbol\delta)
			=
			Q\!\left(
			\sqrt{
				\frac{\beta_{\mathrm c}}{2}
				\boldsymbol\delta^H
				\left(\mathbf G_{\mathrm{co}}^{(K)}\right)^H
				\left(\mathbf R_{\psi}^{(K)}\right)^{-1}
				\mathbf G_{\mathrm{co}}^{(K)}
				\boldsymbol\delta
			}
			\right).
		}
		\label{eq:pep_multiuser}
	\end{equation}
	Thus, the two-user PEP in \eqref{eq:pep_general} is a special case of
	\eqref{eq:pep_multiuser}. The extension only changes the stacked
	cooperative channel and disturbance covariance; the PEP expression
	retains the same form.
	
	\section{PEP-Guided Optimization of the User-Dependent $c_1$}
	\label{sec:c1_optimization}
	
		\textcolor{black}{The PEP expression in Section~\ref{sec:pep_analysis} provides a
		cooperative effective-distance criterion for selecting the user-dependent
		chirp parameters. In this section, we develop an asymmetric $c_1$
		design for the two-user cooperative RSMA-AFDM system. User~1 adopts
		the conventional AFDM parameter that guarantees the required
		single-user path separability, while User~2's $c_1$ is selected by
		minimizing the channel-averaged common-stream PEP. The design therefore
		uses the admissible chirp-parameter degrees of freedom to optimize the
		precoder-dependent cooperative effective distance.}
	
	\subsection{Admissible Set of the $c_1$ Parameter}
	\label{subsec:c1_feasible}
	
	For an integer delay--Doppler channel, the nominal DAFT-domain location
	of the $p$th path of User $k$ is
	\begin{equation}
		d_p^{[k,i]}
		=
		f_p^{[k,i]}+2Nc_{1,k}l_p^{[k,i]},
		\label{eq:c1_path_location}
	\end{equation}
	where $l_p^{[k,i]}$ and $f_p^{[k,i]}$ denote the delay and Doppler
	indices, respectively. 
	
	To preserve the conventional AFDM path-separability condition, we
	adopt the integer chirp-index parameterization
	\begin{equation}
		2Nc_{1,k}=m_k,\qquad m_k\in\mathbb Z.
		\label{eq:integer_chirp_index}
	\end{equation}
	When a fractional-Doppler guard $\xi_f$ is included, the conventional
	minimum value is selected as
	\begin{equation}
		{
			c_{1,\mathrm{ref}}
			=
			\frac{2f_{\max}+2\xi_f+1}{2N}.
		}
		\label{eq:c1_reference}
	\end{equation}
	For the integer-Doppler case, setting $\xi_f=0$ gives
	$c_{1,\mathrm{ref}}=(2f_{\max}+1)/(2N)$.
	
	A conservative no-periodic-aliasing condition over the considered
	delay--Doppler support is
	\begin{equation}
		m_k l_{\max}+2(f_{\max}+\xi_f)<N.
		\label{eq:c1_alias_constraint}
	\end{equation}
	Accordingly, define the admissible integer set
	\begin{equation}
		\begin{aligned}
			\mathcal M
			=&
			\{
			m\in\mathbb Z:
			m\geq 2(f_{\max}+\xi_f)+1,\\&\qquad\qquad\qquad
			ml_{\max}+2(f_{\max}+\xi_f)<N
			\},
		\end{aligned}
		\label{eq:c1_integer_set}
	\end{equation}
	and the corresponding chirp set
	\begin{equation}
		\mathcal C_1
		=
		\left\{
		\frac{m}{2N}:m\in\mathcal M
		\right\}.
		\label{eq:c1_candidate_set}
	\end{equation}
	The lower bound is the conventional single-user AFDM path-separability
	requirement, and the upper bound is used to avoid periodic aliasing
	within the assumed support. These constraints are imposed separately
	for each user. 
	
	\subsection{PEP-Based Optimization Problem}
	\label{subsec:c1_pep_problem}
	
	User~1 is fixed at the conventional reference value
	\begin{equation}
		c_{1,1}=c_{1,\mathrm{ref}}.
		\label{eq:c11_fixed}
	\end{equation}
	Only User~2's $c_1$ chirp parameter is optimized. Specifically, the
	proposed design is formulated as
	\begin{equation}
		{
			c_{1,2}^{\star}
			=
			\arg\min_{c_{1,2}\in\mathcal C_1}
			\overline P_{\mathrm{PEP}}
			\left(
			c_{1,1}=c_{1,\mathrm{ref}},
			c_{1,2}
			\right).
		}
		\label{eq:c1_pep_optimization}
	\end{equation}
	Equivalently, using the integer chirp index,
	\begin{equation}
		{
			m_2^\star
			=
			\arg\min_{m_2\in\mathcal M}
			\overline P_{\mathrm{PEP}}
			\left(
			\frac{2f_{\max}+2\xi_f+1}{2N},
			\frac{m_2}{2N}
			\right),
			c_{1,2}^{\star}=\frac{m_2^\star}{2N}.
		}
		\label{eq:m2_pep_optimization}
	\end{equation}
	
	The objective in \eqref{eq:c1_pep_optimization} is evaluated using the
	conditional PEP in \eqref{eq:pep_general}. For each candidate
	$c_{1,2}$, the user-specific DAFT matrix and equivalent channel are
	reconstructed, and the common-stream precoder is recalculated from
	the resulting channels. 
	
	\subsection{Optimization Procedure}
	\label{subsec:c1_pep_evaluation}
	
	For a finite block length and a practical constellation, exhaustive
	enumeration of all codeword pairs may be computationally expensive.
	The average PEP can therefore be estimated by Monte Carlo averaging
	over channel realizations and a representative set of nonzero
	difference vectors. To avoid missing important error events, the
	difference vectors can be stratified according to their Hamming
	weights. Let $\mathcal D_w$ denote the sampled error-event set with
	Hamming weight $w$. For equiprobable independent QPSK symbols, the
	probability that a randomly selected coordinate is different in an
	ordered symbol pair is $3/4$. Hence, conditioned on a nonzero
	length-$N$ error vector, the Hamming-weight probability is
	\begin{equation}
		\Pr(W=w\mid W>0)
		=
		\frac{
			\binom{N}{w}(3/4)^w(1/4)^{N-w}
		}{
			1-(1/4)^N
		},
	 w=1,\ldots,N.
		\label{eq:qpsk_weight_probability}
	\end{equation}
	A stratified estimate of the average PEP is then
	\begin{equation}
		\begin{aligned}
			\widehat P_{\mathrm{PEP}}(c_{1,2})
			=&
			\frac{1}{N_{\mathrm{ch}}}
			\sum_{q=1}^{N_{\mathrm{ch}}}
			\sum_{w=1}^{N}
			\Pr(W=w\mid W>0)\times
			\\&
			\qquad\frac{1}{|\mathcal D_w|}\sum_{\boldsymbol{\delta}\in\mathcal D_w}
			P_{\mathrm{PEP}}(\boldsymbol{\delta}\mid\mathbf H^{(q)}),
		\end{aligned}
		\label{eq:pep_mc_estimator}
	\end{equation}
	where $N_{\mathrm{ch}}$ is the number of channel realizations. The
	same channel realizations and the same error-event sets should be
	used for every candidate $c_{1,2}$ to obtain a paired and fair
	comparison.

	Since the feasible set $\mathcal C_1$ is discrete and finite, the
	optimization in \eqref{eq:c1_pep_optimization} can be implemented by
	enumeration. The procedure is summarized in Algorithm~\ref{alg:c1_pep}.
	
	\begin{algorithm}[t]
		\caption{PEP-guided optimization of User~2's $c_1$}
		\label{alg:c1_pep}
		\renewcommand{\algorithmicrequire}{\textbf{Input:}}
		\renewcommand{\algorithmicensure}{\textbf{Output:}}
		\begin{algorithmic}[1]
			\REQUIRE $N$, $f_{\max}$, $l_{\max}$, $\xi_f$,
			design SNR, channel statistics,
			and the common-stream precoding rule
			\ENSURE $c_{1,2}^{\star}$
			\STATE Set $c_{1,1}=c_{1,\mathrm{ref}}$ 
			\STATE Construct the admissible set $\mathcal C_1$
			according to \eqref{eq:c1_integer_set}--\eqref{eq:c1_candidate_set}
			\STATE Generate common channel realizations and common
			nonzero error-event samples
			\FOR{each candidate $c_{1,2}\in\mathcal C_1$}
			\STATE Construct the receive-side DAFT matrices
			$\mathbf A_1$ and $\mathbf A_2$
			\STATE Construct the corresponding DAFT-domain
			equivalent channels
			\STATE Recalculate the common-stream precoder
			$\mathbf w_{\mathrm c}$
			\STATE Evaluate the conditional PEP for each channel
			realization and error event
			\STATE Average the PEP according to
			\eqref{eq:pep_mc_estimator}
			\ENDFOR
			\STATE Select the candidate with the minimum estimated
			average PEP
			\STATE Set $c_{1,2}^{\star}=m_2^\star/(2N)$
		\end{algorithmic}
	\end{algorithm}
	
	The selected parameter is finally evaluated using the complete
	RSMA-AFDM receiver, including cooperative common-stream detection,
	successive interference cancellation, and private-stream detection.
	Since the PEP criterion is derived under a Gaussian-disturbance
	coherent-detection model, it is used as a design criterion for the
	EP-based receiver; the resulting BER improvement is established by
	the end-to-end simulations rather than inferred solely from the PEP.

\subsection{Extension to $K\geq 3$ Users}
\label{subsec:c1_multiuser_extension}

For $K\geq3$, User~1 is retained as the reference user with
$c_{1,1}=c_{1,\mathrm{ref}}$, while the remaining users' chirp
parameters are jointly optimized over their AFDM-admissible sets
\begin{equation}
	\mathcal C_{1,k}
	=
	\left\{\frac{m_k}{2N}:m_k\in\mathcal M_k\right\},
	\quad k=2,\ldots,K.
\end{equation}
The optimization problem is
\begin{equation}
	{
		\begin{aligned}
			\min_{\{c_{1,k}\}_{k=2}^{K}}\quad&
			\overline P_{\mathrm{PEP}}^{(K)}
			(c_{1,\mathrm{ref}},c_{1,2},\ldots,c_{1,K})\\
			\mathrm{s.t.}\quad&
			c_{1,k}\in\mathcal C_{1,k},\quad k=2,\ldots,K.
	\end{aligned}}
	\label{eq:c1_multiuser_optimization}
\end{equation}
Here, $\overline P_{\mathrm{PEP}}^{(K)}$ is obtained by averaging the
$K$-user conditional PEP in \eqref{eq:pep_multiuser} over the joint
channel distribution and error events. For each candidate combination,
the user-specific channels and common-stream precoder
$\mathbf w_{\mathrm c}=\mathcal P(\mathbf H_1,\ldots,\mathbf H_K)$
are recalculated. The candidate sets are determined by the
corresponding delay--Doppler supports and need not be identical.
Moreover, the optimized chirp parameters are not required to be
distinct.

The exhaustive search involves $\prod_{k=2}^{K}|\mathcal C_{1,k}|$
candidate combinations. For larger $K$, coordinate-wise discrete
optimization can be adopted, successively updating each user's
$c_1$ while fixing the others. This procedure converges to a
coordinate-wise optimum, although global optimality is not guaranteed.

	\color{black}
	
	\section{Distributed Cooperative EP Detection} 
	\textcolor{black}{This section considers two distributed common-stream detectors. The current decision-fusion EP detector first performs local EP detection and then fuses the local soft decisions using effective-SNR weights. The revised belief-consensus EP detector instead exchanges and sums the external Gaussian likelihood parameters while retaining one shared EP site.}

	\subsection{EP-Based Decision Fusion} 
	Here, we present the decision fusion EP-based detection algorithm for the downlink cooperative RSMA-AFDM system. This method introduces a distributed cooperative mechanism in which the users broadcast their local detection results and fuse the information received from other nodes. Specifically, each user first performs local EP detection on the received signal and then broadcasts the detection result to the other users. Subsequently, all users combine their local EP detection output with the information broadcast by other users, thereby achieving reliable demodulation of the common stream.
	
	The received signal in the DAFT domain for user $k$ can be expressed as 
	\begin{equation}
		\mathbf{y}_k=\sqrt{\beta_\mathrm{c}}\mathbf{G}_{\mathrm{c},k}\mathbf{X}_\mathrm{c}+{\boldsymbol{\psi}_k},
		\label{io1g}
	\end{equation}
	where $\mathbf{G}_{\mathrm{c},k}=\mathbf{H}_k\mathbf{w}_\mathrm{c}\in\mathbb{C}^{N\times N}$ is the equivalent channel of User $k$.
	
	The \textit{a posteriori} distribution of the common-stream symbols is
	approximated as
	\begin{equation}
		p(\mathbf{X}_{\mathrm c}|\mathbf{y}_k)
		\propto
		\mathcal{CN}\!\left(
		\mathbf{y}_k;
		\sqrt{\beta_{\mathrm c}}\mathbf{G}_{\mathrm c,k}
		\mathbf{X}_{\mathrm c},
		\mathbf{R}_{\psi,k}
		\right)
		\prod_{m=1}^{N}p(X_{\mathrm c}[m]),
	\end{equation}
	where \begin{equation}
		\mathbf{R}_{\psi,k}
		=
		\sum_{v=1}^{K}
		\beta_{\mathrm p,v}
		\mathbf{H}_{k}\mathbf{w}_{\mathrm p,v}
		\mathbf{w}_{\mathrm p,v}^{H}\mathbf{H}_{k}^{H}
		+
		\sigma_{k}^{2}\mathbf{I}_{N},\end{equation} and $p(\mathbf{X}_\mathrm{c})$ represents the \textit{a priori} distribution of $\mathbf{X}_\mathrm{c}[m]$, which is denoted as
	\begin{equation}
		p(\mathbf{X}_\mathrm{c})=\prod_{m=1}^{N}\mathbb{I}({X}_\mathrm{c}[m]\in\mathcal{A}).
		\label{pd}
	\end{equation}

	Since $p(\mathbf{X}_\mathrm{c})$ is not Gaussian, computing $p(\mathbf{X}_\mathrm{c}|\mathbf{y}_k) $ directly would require traversing all possible values of $\mathbf{X}_\mathrm{c}$, which results in exponential complexity and makes practical implementation prohibitive \cite{10845213}. Therefore, we employ the EP algorithm to approximate the solution to $p(\mathbf{X}_\mathrm{c}|\mathbf{y}_k)$. The EP detection algorithm provides approximate ML solutions while significantly reducing the computational complexity. EP constructs a simple Gaussian distribution $q(\mathbf{X}_\mathrm{c})$ to approximate $p(\mathbf{X}_\mathrm{c}|\mathbf{y}_k)$. Through moment matching, EP enables the statistical properties of the constructed $q(\mathbf{X}_\mathrm{c})$ closely align with those of $p(\mathbf{X}_\mathrm{c})$,  thereby supporting detection performance.
	
	The received symbol vector $\mathbf{y}_k$, as shown in (\ref{io1g}), can be transformed into a real-valued symbol representation as
	\begin{equation}
		\widetilde{\mathbf{y}}_k =\sqrt{\beta_\mathrm{c}}\widetilde{\mathbf{G}}_{\mathrm{c},k}\widetilde{\mathbf{X}}_\mathrm{c}+{\boldsymbol{\widetilde{\psi}}}_k,
		\label{ior}
	\end{equation}
	where 
	\begin{equation}
		\begin{aligned}
			\widetilde{\mathbf{y}}_k =&[\mathcal{R}\{\mathbf{y}_k\}^{T}\quad\mathcal{I}\{\mathbf{y}_k\}^{T}]^{T}\in\mathbb{R}^{2N\times1}, \\
			\widetilde{\mathbf{X}}_\mathrm{c} =&[\mathcal{R}\{{\mathbf{X}}_\mathrm{c}\}^{T}\quad\mathcal{I}\{{\mathbf{X}}_\mathrm{c}\}^{T}]^{T}\in\mathbb{R}^{2N\times1}, \\
			{\boldsymbol{\widetilde{\psi}}}_k =&[\mathcal{R}\{{\boldsymbol{{\psi}}}_k\}^{T}\quad\mathcal{I}\{{\boldsymbol{{\psi}}}_k\}^{T}]^{T}\in\mathbb{R}^{2N\times1}, \\
			\widetilde{\mathbf{G}}_{\mathrm{c},k} =&
			\begin{bmatrix}
				\mathcal R\{\mathbf G_{\mathrm c,k}\}
				&
				-\mathcal I\{\mathbf G_{\mathrm c,k}\}\\
				\mathcal I\{\mathbf G_{\mathrm c,k}\}
				&
				\mathcal R\{\mathbf G_{\mathrm c,k}\}
			\end{bmatrix}\in\mathbb{R}^{2N\times2N}, 
		\end{aligned}
	\end{equation}
	where $\mathbb{R}$ signifies the set of real numbers; $\mathcal{R}{(\mathbf{a})}$ and $\mathcal{I}{(\mathbf{a})}$ denote the real and imaginary parts, respectively.
	
	Then $q(\widetilde{\mathbf{X}}_\mathrm{c})=\prod_{n=1}^{2N}e^{\gamma_n^{(l)}\widetilde{{X}}_\mathrm{c}[n]-\frac{1}{2}\lambda_n^{(l)}\widetilde{{X}}_\mathrm{c}[n]^2}$ is constructed to replace the original \textit{a priori} distribution $p(\widetilde{\mathbf{X}}_\mathrm{c})$, where $\gamma_n^{(l)}$ and $\lambda_n^{(l)}$ denote two parameters that are iteratively updated to achieve a more accurate approximation of the true distribution; $l$ is the EP iteration index.
	
	Both $p(\widetilde{\mathbf{y}}_k|{\widetilde{\mathbf{X}}}_\mathrm{c})$ and $q^{(l)}({\widetilde{\mathbf{X}}}_\mathrm{c})$ follow Gaussian distributions, such that their product $p^{(l)}({\widetilde{\mathbf{X}}}_\mathrm{c}|\widetilde{\mathbf{y}}_k)$ also follows a Gaussian distribution. Through theoretical derivation, we have $p^{(l)}({\widetilde{\mathbf{X}}}_\mathrm{c}|\widetilde{\mathbf{y}}_k)\propto\mathcal{N}(\widetilde{\mathbf{X}}_\mathrm{c}:\boldsymbol{\mu}^{(l)},\boldsymbol{\Sigma}^{(l)})$, whose explicit expression is given by
	\begin{equation}
		\begin{aligned}
			\boldsymbol{\Sigma}^{(l)}
			={}&
			\left(
			2\beta_{\mathrm c}
			\widetilde{\mathbf{G}}_{\mathrm c,k}^{T}
			\widetilde{\mathbf{R}}_{\psi,k}^{-1}
			\widetilde{\mathbf{G}}_{\mathrm c,k}
			+
			\operatorname{diag}
			\{\boldsymbol{\lambda}^{(l-1)}\}
			\right)^{-1},\\
			\boldsymbol{\mu}^{(l)}
			={}&
			\boldsymbol{\Sigma}^{(l)}
			\left(
			2\sqrt{\beta_{\mathrm c}}\,
			\widetilde{\mathbf{G}}_{\mathrm c,k}^{T}
			\widetilde{\mathbf{R}}_{\psi,k}^{-1}
			\widetilde{\mathbf{y}}_k
			+
			\boldsymbol{\gamma}^{(l-1)}
			\right).
		\end{aligned}
		\label{mv}
	\end{equation}
	where $\boldsymbol{\lambda}^{(l-1)}=[\lambda_{1}^{(l-1)},\ldots,\lambda_{2N}^{(l-1)}]^{T}\in\mathbb{R}^{2N\times1}$; $\boldsymbol{\mu}^{(l)}= [\mu_{1}^{(l)},\ldots,\mu_{2N}^{(l)}]^{T}\in\mathbb{R}^{2N\times1}$;  $\boldsymbol{\gamma}^{(l-1)}= [\gamma_{1}^{(l-1)},\ldots,\gamma_{2N}^{(l-1)}]^{T}\in\mathbb{R}^{2N\times1}$; $\Sigma_n^{(l)}=[\boldsymbol{\Sigma}^{(l)}]_{n,n}$
	denotes the $n$th marginal posterior variance; $\widetilde{\mathbf{R}}_{\psi,k}
	=
	\begin{bmatrix}
		\mathcal{R}\{\mathbf{R}_{\psi,k}\}
		&
		-\mathcal{I}\{\mathbf{R}_{\psi,k}\}\\
		\mathcal{I}\{\mathbf{R}_{\psi,k}\}
		&
		\mathcal{R}\{\mathbf{R}_{\psi,k}\}
	\end{bmatrix},$
	
	Given the independence of transmitted symbols, the $n$-th marginal of the Gaussian approximation is
	\begin{equation}
		p_n^{(l)}(\widetilde X_{\mathrm c}[n]|
		\widetilde{\mathbf y}_k)
		=
		\mathcal N\!\left(
		\widetilde X_{\mathrm c}[n];
		\mu_n^{(l)},\Sigma_n^{(l)}
		\right),
	\end{equation}
	where
	$\Sigma_n^{(l)}=[\boldsymbol\Sigma^{(l)}]_{n,n}$.
	Next, the DAFT-domain cavity distribution \cite{122} is obtained by dividing $p^{(l)}(\widetilde{{X}}_\mathrm{c}[n]|\widetilde{\mathbf{y}}_k)$ by $q^{(l-1)}(\widetilde{{X}}_\mathrm{c}[n])$  as
	\begin{equation}
		p^{{(l)}\setminus n}(\widetilde{{X}}_\mathrm{c}[n]|\widetilde{\mathbf{y}}_k)\triangleq\frac{p^{(l)}(\widetilde{{X}}_\mathrm{c}[n]|\widetilde{\mathbf{y}}_k)}{q^{(l-1)}(\widetilde{{X}}_\mathrm{c}[n])}\propto\mathcal{N}\left(\widetilde{{X}}_\mathrm{c}[n]:E_n^{(l)},{\varepsilon}_n^{(l)}\right),
		\label{ca}
	\end{equation}
	where 
	\begin{equation}
		\begin{aligned}
			&{\varepsilon}_n^{(l)} =\frac{\Sigma_{n}^{(l)}}{1-\Sigma_{n}^{(l)}\lambda_{n}^{(l-1)}}, \\
			&E_n^{(l)} ={\varepsilon}_n^{(l)}\left(\frac{\mu_n^{(l)}}{\Sigma_n^{(l)}}-\gamma_n^{(l-1)}\right).
		\end{aligned}
		\label{obs}
	\end{equation}
	
	Then, it is necessary to calculate the moments of $p^{{(l)}\setminus n}(\widetilde{{X}}_\mathrm{c}[n]|\widetilde{\mathbf{y}}_k)p(\widetilde{{X}}_\mathrm{c}[n])$. Given the random nature of the transmitted signal, $p(\widetilde{{X}}_\mathrm{c}[n])$ may be regarded as a uniform distribution over the space of constellation points. The first-order moment $\chi_{n}^{(l)}$ and second-order moment ${V}_{n}^{(l)}$ can be computed as
	\begin{equation}
		\begin{aligned}&\chi_{n}^{(l)}=\sum_{a\in\widetilde{\mathcal{A}}} a\times q^{{(l)}\setminus n}(\widetilde{{X}}_\mathrm{c}[n]=a),\\&{V}_{n}^{(l)}=\sum_{a\in\widetilde{\mathcal{A}}} (a-\chi_{n}^{(l)})^{2}\times q^{{(l)}\setminus n}(\widetilde{{X}}_\mathrm{c}[n]=a).
		\end{aligned}
		\label{pj}
	\end{equation}
	where $\widetilde{\mathcal{A}}$ is the set of real-valued constellation points.
	Then, moment matching is performed between $p^{{(l)}\setminus n}(\widetilde{{X}}_\mathrm{c}[n]|\widetilde{\mathbf{y}}_k)p(\widetilde{{X}}_\mathrm{c}[n])$ and $p^{{(l)}\setminus n}(\widetilde{{X}}_\mathrm{c}[n]|\widetilde{\mathbf{y}}_k)q^{(l)}(\widetilde{{X}}_\mathrm{c}[n])$ to update the parameters of $q^{(l)}(\widetilde{{X}}_\mathrm{c}[n])$ as
	\begin{equation}
		\begin{aligned}&\widehat{\boldsymbol{\lambda}}^{(l)}=(\mathbf{V}^{(l)})^{-1}-(\boldsymbol{\varepsilon}^{(l)})^{-1},\\&\widehat{\boldsymbol{\gamma}}^{(l)}=(\mathbf{V}^{(l)})^{-1}\boldsymbol{\chi}^{(l)}-(\boldsymbol{\varepsilon}^{(l)})^{-1}\mathbf{E}^{(l)},
		\end{aligned}
		\label{update}
	\end{equation}
	where ${\mathbf{V}}^{(l)}=\operatorname{diag}([V_{1}^{(l)},\ldots,V_{2N}^{(l)}])\in\mathbb{R}^{2N\times 2N}$; $ \boldsymbol{\chi}^{(l)}=[{\chi}_{1}^{(l)},\ldots,{\chi}_{2N}^{(l)}]^T\in\mathbb{R}^{2N\times1}$; $\boldsymbol{\varepsilon}^{(l)}= \operatorname{diag}([\varepsilon^{(l)}_{1},\ldots,\varepsilon^{(l)}_{2N}])\in\mathbb{R}^{2N\times 2N}$; $\mathbf{E}^{(l)}=  [E^{(l)}_{1},\ldots,E^{(l)}_{2N}]^{{T}}\in\mathbb{R}^{2N\times1}$.
	
	Finally, a damping parameter $\eta$ is employed to control the convergence rate, which can be expressed as
	\begin{equation}
		\begin{aligned}
			\boldsymbol{\lambda}^{(l)}
			&=
			(1-\eta)\boldsymbol{\lambda}^{(l-1)}
			+\eta\widehat{\boldsymbol{\lambda}}^{(l)},\\
			\boldsymbol{\gamma}^{(l)}
			&=
			(1-\eta)\boldsymbol{\gamma}^{(l-1)}
			+\eta\widehat{\boldsymbol{\gamma}}^{(l)},
		\end{aligned}
		\label{mm}
	\end{equation}
	
	The above steps are iterated until either the preset maximum number of iterations $L$ is reached or the target accuracy is achieved. 
	
	After each user performs EP detection on the common data stream locally, it broadcasts the detection result $\boldsymbol{\chi}_k$ to other users. Subsequently, all users fuse the local EP detection output expectation $\boldsymbol{\chi}_k$ with the cooperative information $\boldsymbol{\chi}_v (v = 1, \ldots, K, v \neq k)$ received from neighboring users to achieve a consistent estimate of the global message. Specifically, under the assumption that the channel gains are normalized, this effective-SNR-based reliability weighting assigns larger weights
	to users with stronger post-whitening common-stream channels, thereby
	improving the reliability of the fused estimate, which can be expressed as
	\begin{equation}
		\boldsymbol{\chi}_{\mathrm{co}}^{(L)}
		=
		\sum_{v=1}^{K}
		\omega_v
		\boldsymbol{\chi}_{v}^{(L)}.
	\end{equation}
	where
	\begin{equation}
		\omega_v
		=
		\frac{\xi_v}{\sum_{j=1}^{K}\xi_j},	\qquad
		\xi_v
		=
		\frac{1}{N}
		\operatorname{tr}\!\left(
		\mathbf{G}_{\mathrm c,v}^{H}
		\mathbf{R}_{\psi,v}^{-1}
		\mathbf{G}_{\mathrm c,v}
		\right).
		\label{eq:fusion_weight}
	\end{equation}
	
	\subsection{EP-Based Belief Consensus}
	\textcolor{black}{The decision-fusion detector combines the final local EP outputs only. To exploit the observation information generated at every node without counting the shared EP site repeatedly, the belief-consensus update is revised to exchange and sum the external natural parameters of the local Gaussian likelihoods.}
	
	\textcolor{black}{For user $v$, define the external likelihood precision matrix $\mathbf J_v\in\mathbb{R}^{2N\times2N}$ and natural information vector $\mathbf b_v\in\mathbb{R}^{2N\times1}$, excluding the common EP site, as}
	{\color{black}
	\begin{equation}
	\begin{aligned}
	\mathbf J_v&=2\beta_{\mathrm c}\widetilde{\mathbf G}_{\mathrm c,v}^{T}
	\widetilde{\mathbf R}_{\psi,v}^{-1}\widetilde{\mathbf G}_{\mathrm c,v},\\
	\mathbf b_v&=2\sqrt{\beta_{\mathrm c}}\widetilde{\mathbf G}_{\mathrm c,v}^{T}
	\widetilde{\mathbf R}_{\psi,v}^{-1}\widetilde{\mathbf y}_v .
	\end{aligned}
	\label{eq:external_info}
	\end{equation}}
	\textcolor{black}{With the subscript $\mathrm{co}$ denoting the cooperative update, the external information is summed across users and combined with one shared EP site:}
	{\color{black}
	\begin{equation}
	\begin{aligned}
	\mathbf J_{\mathrm{co}}&=\sum_{v=1}^{K}\mathbf J_v,\qquad
	\mathbf b_{\mathrm{co}}=\sum_{v=1}^{K}\mathbf b_v,\\
	\boldsymbol\Sigma_{\mathrm{co}}^{(l)}&=\left(\mathbf J_{\mathrm{co}}+\operatorname{diag}\{\boldsymbol\lambda^{(l-1)}\}\right)^{-1},\\
	\boldsymbol\mu_{\mathrm{co}}^{(l)}&=\boldsymbol\Sigma_{\mathrm{co}}^{(l)}
	\left(\mathbf b_{\mathrm{co}}+\boldsymbol\gamma^{(l-1)}\right).
	\end{aligned}
	\label{coup}
	\end{equation}}
	\textcolor{black}{Here, $\boldsymbol\Sigma_{\mathrm{co}}^{(l)}$ and $\boldsymbol\mu_{\mathrm{co}}^{(l)}$ are the cooperative Gaussian posterior covariance and mean; $\boldsymbol\lambda^{(l)}$ and $\boldsymbol\gamma^{(l)}$ remain the shared EP-site parameters defined in Subsection V-A. Their cavity, moment-matching, and damping updates follow \eqref{obs}--\eqref{mm}, with $\Sigma_n^{(l)}$ and $\mu_n^{(l)}$ replaced by the corresponding entries of $\boldsymbol\Sigma_{\mathrm{co}}^{(l)}$ and $\boldsymbol\mu_{\mathrm{co}}^{(l)}$.}
	
	\begin{algorithm}[t]
		\caption{The belief consensus-based distributed cooperative detection for common stream in RSMA-AFDM.}
		\renewcommand{\algorithmicrequire}{\textbf{Input: }}
		\renewcommand{\algorithmicensure}{\textbf{Initial:}}
		\begin{algorithmic}[1]
			\REQUIRE   \textcolor{black}{At User $v$: $\widetilde{\mathbf{y}}_v$, $\widetilde{\mathbf{G}}_{\mathrm c,v}$, $\widetilde{\mathbf R}_{\psi,v}$; shared mean symbol energy $E_{s}$, $L$, $\eta$.} 
			\ENSURE $\gamma_{n}=0, \lambda_{n}=E_{s}^{-1}$.    
			\STATE \textcolor{black}{Each user forms $(\mathbf J_v,\mathbf b_v)$ by (\ref{eq:external_info}) and exchanges them to obtain $(\mathbf J_{\mathrm{co}},\mathbf b_{\mathrm{co}})$;}
			\FOR{$l$ = $1:L$} 
			\STATE \textcolor{black}{Compute $\boldsymbol\Sigma_{\mathrm{co}}^{(l)}$ and $\boldsymbol\mu_{\mathrm{co}}^{(l)}$ by (\ref{coup});}
			\STATE Compute ${\varepsilon}_n^{(l)}$ and $E_n^{(l)}$ by (\ref{obs});
			\STATE Compute $\chi_{n}^{(l)}$ and ${V}_{n}^{(l)}$ by (\ref{pj});	
			\STATE Compute $\lambda_n^{(l)}$ and $\gamma_n^{(l)}$ by (\ref{update});
			\STATE Update $\lambda_n^{(l)}$ and $\gamma_n^{(l)}$ by (\ref{mm});		
			\STATE \textcolor{black}{Each user retains the updated shared-site parameters $\boldsymbol\lambda^{(l)}$ and $\boldsymbol\gamma^{(l)}$ locally;}
			\ENDFOR
			\STATE Eliminate the common stream component by (\ref{elicom});
			\STATE Detect the private stream by (\ref{mmse});
			\RETURN $\hat{\mathbf{X}}$.		
		\end{algorithmic}
		\label{al}
	\end{algorithm}
	
	\textcolor{black}{Because the exchanged quantities are external likelihood parameters, their sum incorporates all users' observations while retaining a single shared EP site. This avoids the repeated counting of a common prior and provides the Gaussian-information form used in the output-SNR analysis below.}
	
	After detecting the common stream, its component is removed from the received signal $\mathbf{y}_k$ as
	\begin{equation}
		\begin{aligned}
			\hat{\mathbf{y}}_{\mathrm{p},k}&=\mathbf{y}_k-\sqrt{\beta_\mathrm{c}}{\mathbf{G}}_{\mathrm{c},k}\hat{\mathbf{X}}_\mathrm{c},
		\end{aligned}
		\label{elicom}
	\end{equation}
	where $\hat{\mathbf{X}}_\mathrm{c}$ represents the common stream estimated by the proposed cooperative detection scheme; $\hat{\mathbf{y}}_{\mathrm{p},k}$ denotes the private stream component obtained after SIC. Assuming perfect common stream estimation, $\hat{\mathbf{y}}_{\mathrm{p},k}$ is expressed as
	\begin{equation}
		\hat{\mathbf{y}}_{\mathrm{p},k}=\mathbf{H}_k\sum_{v=1}^{K}\sqrt{\beta_\mathrm{p,v}}\mathbf{w}_{\mathrm{p},v}\mathbf{X}_{\mathrm{p},v}+\mathbf{z}_k.
	\end{equation}
	\textcolor{black}{Under perfect CSI and the stated full-row-rank condition, let
	$\alpha=\|\mathbf H^H(\mathbf H\mathbf H^H)^{-1}\|_{\mathrm F}^{-1}$.
	The ZF formula gives $\mathbf H\mathbf D=\alpha\mathbf I_{KN}$ and hence
	$\mathbf H_k\mathbf w_{\mathrm{p},v}=\alpha\delta_{kv}\mathbf I_N$.
	This identity also holds for correlated user channels when the stacked channel retains full row rank. Thus, $\hat{\mathbf{y}}_{\mathrm{p},k}$ becomes
	\begin{equation}
		\hat{\mathbf{y}}_{\mathrm{p},k}=\sqrt{\beta_\mathrm{p,k}}\mathbf{H}_k\mathbf{w}_{\mathrm{p},k}\mathbf{X}_{\mathrm{p},k}+\mathbf{z}_k = \sqrt{\beta_\mathrm{p,k}}\mathbf{G}_{\mathrm{p},k}\mathbf{X}_{\mathrm{p},k}+\mathbf{z}_k,
	\end{equation}
	where $\mathbf{G}_{\mathrm{p},k}=\mathbf H_k\mathbf w_{\mathrm{p},k}=\alpha\mathbf I_N$ is the local private-stream channel under these assumptions.} The private stream is detected using minimum mean square error (MMSE), yielding
	\begin{equation}
		\hat{\mathbf{X}}_{\mathrm{p},k}=(\mathbf{G}_{\mathrm{p},k}^{H} \mathbf{G}_{\mathrm{p},k}+\frac{\sigma_k^2}{{\beta_\mathrm{p,k}}}\mathbf{I}_{N})^{-1} \mathbf{G}_{\mathrm{p},k}^{H}\hat{\mathbf{y}}_{\mathrm{p},k}/\sqrt{\beta_\mathrm{p,k}}.
		\label{mmse}
	\end{equation}
	The detailed process of the proposed cooperative RSMA-AFDM detection is described in Algorithm \ref{al}.

	\subsection{\textcolor{black}{Output-SNR Analysis of External-Information Fusion}}
	\label{subsec:output_snr}
	{\color{black}
	Let $\boldsymbol\Sigma_k^{(l)}$ denote the local Gaussian posterior covariance in \eqref{mv} for user $k$, evaluated with the same EP-site precision $\boldsymbol\lambda^{(l-1)}$ as the cooperative update. Since $\widetilde{\mathbf R}_{\psi,v}\succ0$ implies $\mathbf J_v\succeq0$, the cooperative precision satisfies
	\begin{equation}
	\begin{aligned}
	\left(\boldsymbol\Sigma_{\mathrm{co}}^{(l)}\right)^{-1}
	&=\operatorname{diag}\{\boldsymbol\lambda^{(l-1)}\}
	+\sum_{v=1}^{K}\mathbf J_v\\
	&\succeq\operatorname{diag}\{\boldsymbol\lambda^{(l-1)}\}+\mathbf J_k
	=\left(\boldsymbol\Sigma_k^{(l)}\right)^{-1}.
	\end{aligned}
	\label{eq:snr_info_sum}
	\end{equation}
	When these precision matrices are positive definite, $\boldsymbol\Sigma_{\mathrm{co}}^{(l)}\preceq\boldsymbol\Sigma_k^{(l)}$. Thus, for the $n$th real common-stream symbol defined in \eqref{ior}, the Gaussian-EP variance-based output SNR obeys
	\begin{equation}
	\operatorname{SNR}_{\mathrm{co},n}^{(l)}
	\triangleq\frac{\mathbb E[\widetilde X_{\mathrm c}[n]^2]}
	{[\boldsymbol\Sigma_{\mathrm{co}}^{(l)}]_{n,n}}
	\geq\frac{\mathbb E[\widetilde X_{\mathrm c}[n]^2]}
	{[\boldsymbol\Sigma_k^{(l)}]_{n,n}}
	\triangleq\operatorname{SNR}_{k,n}^{(l)},
	\label{eq:snr_ep}
	\end{equation}
	where $[\cdot]_{n,n}$ denotes the $n$th diagonal entry and the shared numerator is the real-symbol power. This bound concerns the Gaussian EP approximation. For the decision-fusion rule, let $\nu_v=(2N)^{-1}\mathbb E\|\boldsymbol\chi_v^{(L)}-\widetilde{\mathbf X}_{\mathrm c}\|^2$ be the local soft-decision error variance per real component. With independent, zero-mean, isotropic local errors, the fused variance is $\nu_{\mathrm{co}}=\sum_v\omega_v^2\nu_v$. Minimizing it under $\sum_v\omega_v=1$ yields $\omega_v^{\star}=\nu_v^{-1}/\sum_j\nu_j^{-1}$, which maximizes the linear-fusion output SNR for fixed symbol power. Hence, the effective-SNR weights in \eqref{eq:fusion_weight} approximate this optimum when $\xi_v$ tracks $\nu_v^{-1}$.}
	
	\section{Simulation Results and Complexity Analysis}
\color{black}
\subsection{Numerical Results}
	\textcolor{black}{In this subsection, we evaluate the proposed PEP-guided cooperative effective-distance optimization and the distributed detectors through simulations.} We also evaluate the proposed decision-fusion cooperative detection scheme and the belief consensus-based cooperative EP detection scheme. MMSE, EP \cite{11263921}, and CD-MAMP \cite{Chi2026CDMAMP} are used as benchmarks. We consider a downlink RSMA-AFDM system with two users. The power allocation factor for the common stream is $\beta_\mathrm{c}=0.75$. The number of symbols per AFDM frame is set to $N=64$, the number of paths is $P=4$, with maximum delay index \(l_{\mathrm{max}}=2\) and maximum Doppler index \(f_{\mathrm{max}}=2\), where both delay and Doppler indices are fractional. The chirp parameter $c_{1,1}$ for User 1 is set to $\frac{2f_\mathrm{max}+2\xi_{f}+1}{2N}$, where $\xi_{f}=1$. The proposed joint PEP-guided design selects $c_{1,2}=\frac{18}{2N}$ for User 2. Furthermore, both users' $c_2$ is set to $\frac{1}{N\pi}$.
	
	\begin{figure*}[t]
		\centering
		\begin{minipage}[c]{0.32\linewidth} 
			\centering
			\includegraphics[width=1\textwidth]{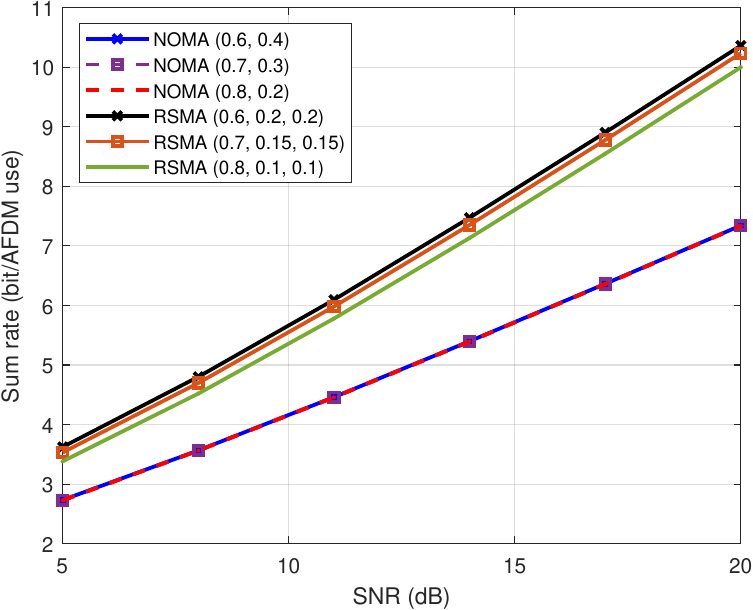}
			\caption{\textcolor{black}{Sum rate for RSMA-AFDM and NOMA-AFDM under different power allocations.}}
			\label{ma}
		\end{minipage}
		\begin{minipage}[c]{0.32\linewidth} 
			\centering
			\includegraphics[width=1\textwidth]{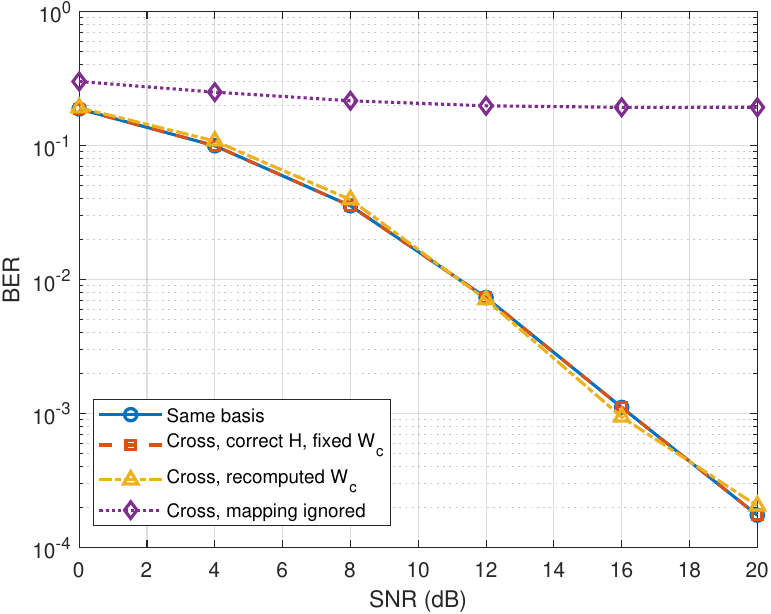}
			\caption{\textcolor{black}{Common-stream BER for same-basis and cross-basis detection.}}
			\label{cb}
		\end{minipage}
		\begin{minipage}[c]{0.32\linewidth}
			\centering
			\includegraphics[width=1\textwidth]{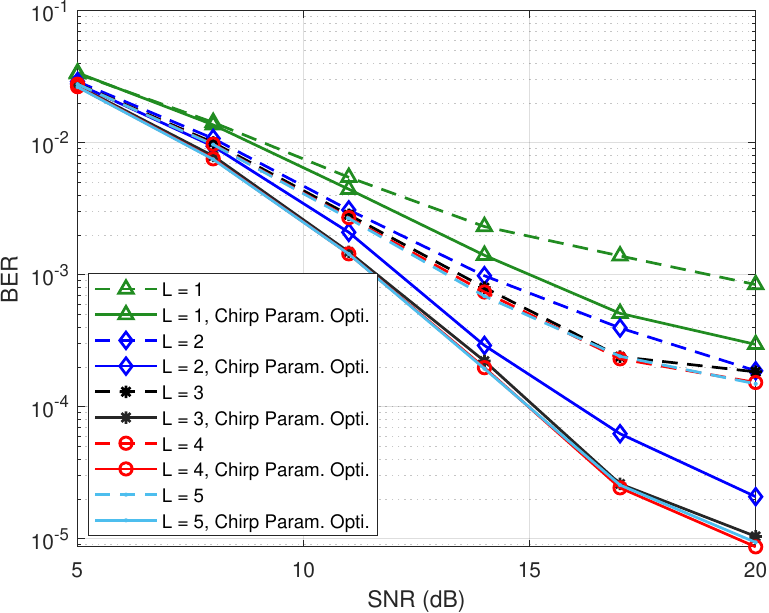}
			\caption{\textcolor{black}{Common-stream BER for belief consensus-based scheme with different iteration counts $L$.}}
			\label{inte}
		\end{minipage}
	\end{figure*}
	
First, Fig. \ref{ma} compares the achievable sum rates of RSMA-AFDM with $(\beta_\mathrm{c},\beta_{p1},\beta_{p2})\in\{(0.6,0.2,0.2)$, $(0.7,0.15,0.15)$, $(0.8,0.1,0.1)\}$ and shared-precoder PD-NOMA-AFDM with $\{(0.6,0.4),(0.7,0.3),(0.8,0.2)\}$ for $N_t=8$ and $K=2$ under Gaussian signaling and ideal SIC. Across all SNRs and power splits, RSMA-AFDM consistently outperforms PD-NOMA-AFDM, as separately precoding the RSMA private streams mitigates inter-user interference more effectively than the shared-precoder NOMA baseline. Moreover, assigning more power to the RSMA private streams increases the sum rate because private-stream gains outweigh the common-stream loss, whereas the PD-NOMA sum rate remains nearly invariant across power splits due to mutually offsetting user rates
	
Fig. \ref{cb} evaluates the impact of receive-side DAFT basis disparity via $\mathbf{C}_{21} = \mathbf{A}_2 \mathbf{A}_1^H$ across four setups: 1) Same basis (identical chirp parameters); 2) Cross, correct $\mathbf{H}$, fixed $\mathbf{W}_c$ (User 2 accounts for $\mathbf{C}_{21}$ in the effective channel and disturbance covariance); 3) Cross, recomputed $\mathbf{W}_c$ (precoder redesigned with embedded $\mathbf{C}_{21}$); and 4) Cross, mapping ignored. As shown, properly incorporating the unitary mapping $\mathbf{C}_{21}$—with either fixed or recomputed $\mathbf{W}_c$—yields common-stream BER curves that virtually overlap with the Same basis benchmark, confirming negligible residual loss due to unitary invariance. Conversely, ignoring $\mathbf{C}_{21}$ causes severe model mismatch and an irreducible BER floor around $0.2$.
	
Next, we investigate the impact of the iteration number $L$ on the proposed belief consensus EP-based distributed  cooperative detector. Fig. \ref{inte} shows the simulated BER performance for different values of $L$. For each detector, the suffix "Chirp Param. Opti." indicates that the user-dependent parameters are optimized by minimizing the channel-averaged PEP; otherwise, the conventional admissible chirp parameters are employed. The results show that the algorithm converges rapidly as $L$ increases: after only three iterations, the performance improves significantly and is already close to the optimum; when $L=4$, the performance curve becomes almost flat, and further increasing $L$ yields negligible additional gain. This rapid convergence characteristic indicates that the proposed belief consensus EP-based cooperative detection algorithm can effectively exploit cooperative information. Based on this observation, we set $L = 4$ in all subsequent simulations to achieve a balance between detection performance and computational complexity.	
	
\begin{figure}
	\centering
	\includegraphics[width=0.34\textwidth]{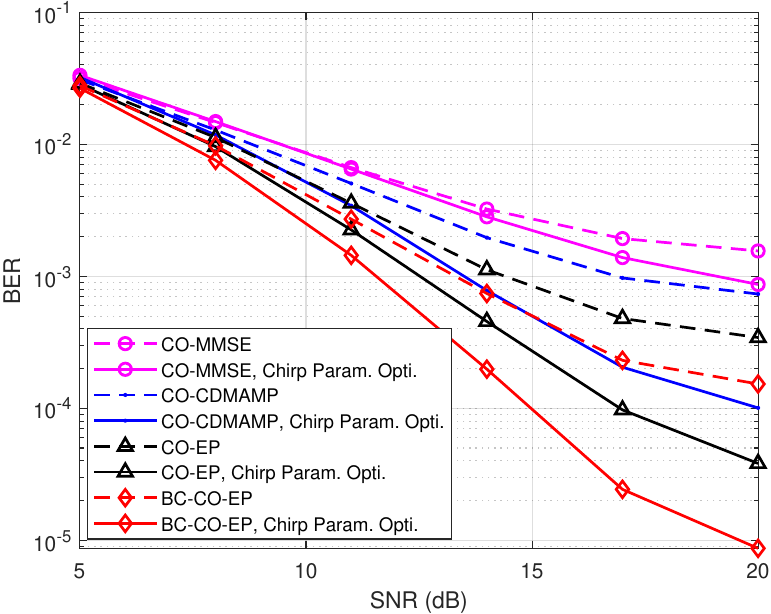}
	\caption{\textcolor{black}{Common-stream BER for different detection schemes with and without chirp-parameter optimization ($N_t=4$, $K=2$).}}
	\label{compcpode}
\end{figure}	
	
	\begin{figure*}[t]
		\centering
		\subfigure[Common-stream BER]{
			\begin{minipage}[c]{0.31\linewidth} 
				\centering
				\includegraphics[width=1.0\linewidth]{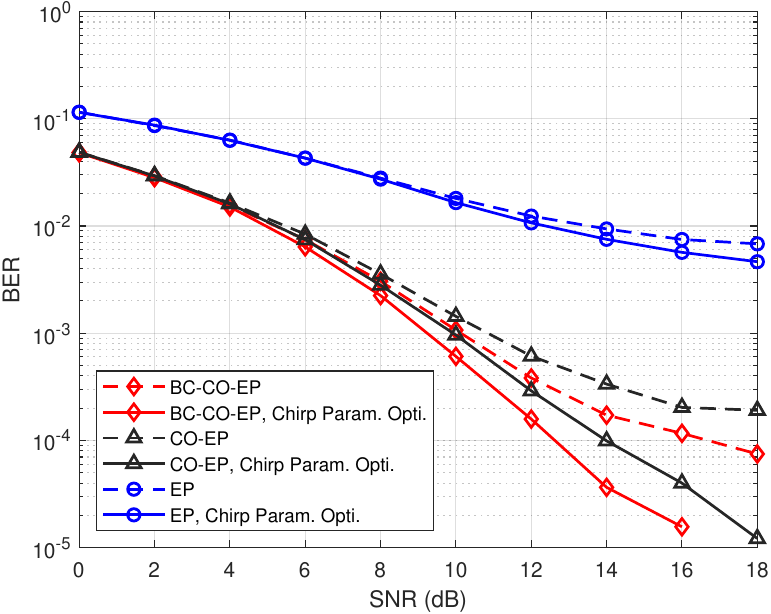}\vspace{-0.0pt}  
				\label{Nt8_co}
			\end{minipage}
		}
		\subfigure[Private-stream BER]{
			\begin{minipage}[c]{0.31\linewidth}
				\centering
				\includegraphics[width=1.0\linewidth]{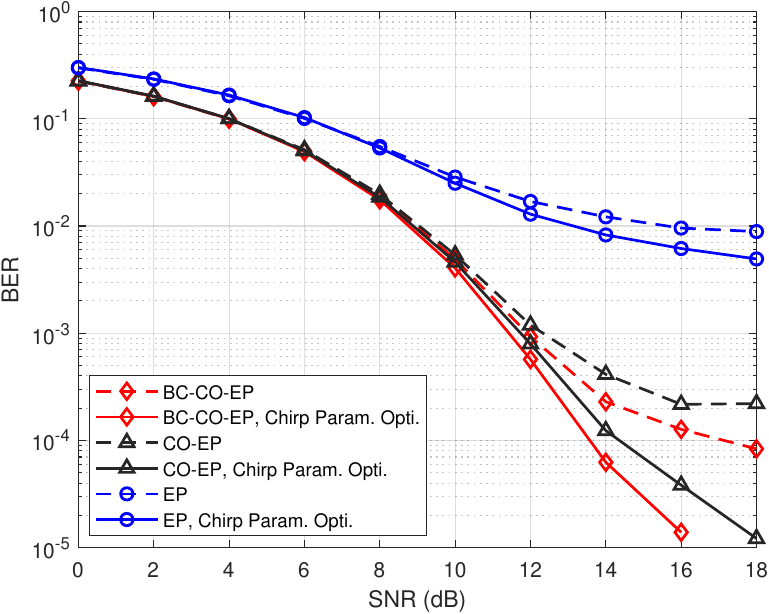}\vspace{0pt}
				\label{Nt8pr}
			\end{minipage}
		}
		\subfigure[Overall BER]{
			\begin{minipage}[c]{0.31\linewidth}
				\centering
				\includegraphics[width=1.0\linewidth]{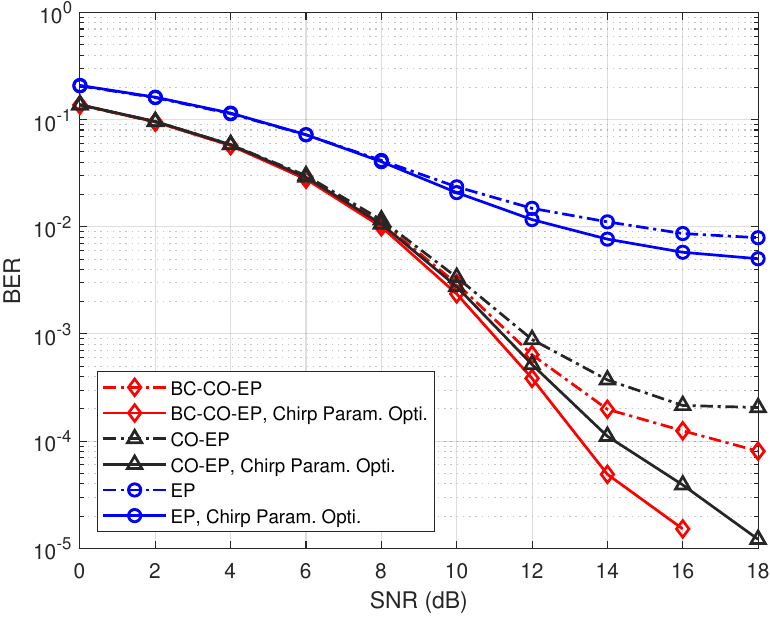}\vspace{0pt}
				\label{Nt8}
			\end{minipage}
		}
		\caption{\textcolor{black}{Common-stream, private-stream, and overall BER for different detection schemes with and without chirp-parameter optimization ($N_t=8$, $K=2$).}}
		\label{PNt8}
	\end{figure*}

\begin{figure*}[t]
	\centering
	\begin{minipage}[c]{0.32\linewidth}
		\centering
		\includegraphics[width=1\textwidth]{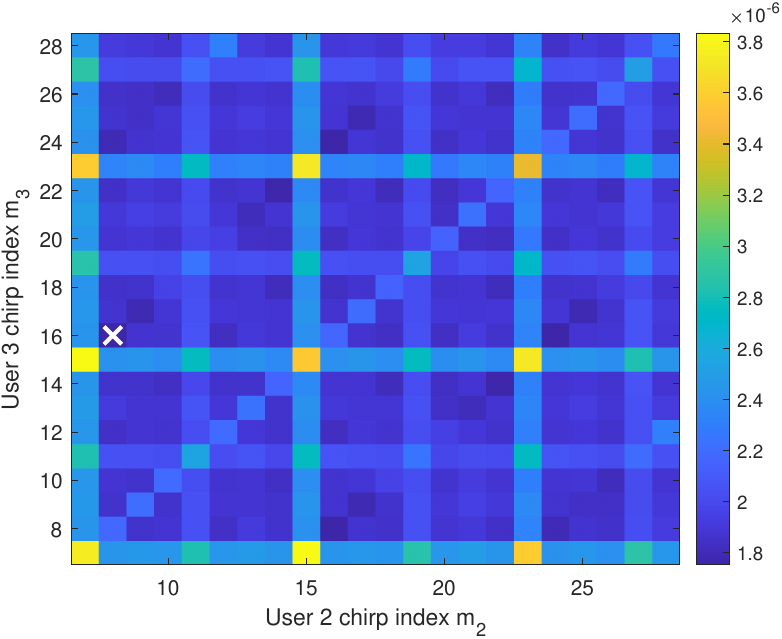}
		\caption{\textcolor{black}{Average PEP objective for three-user chirp-parameter optimization.}}
		\label{3userc1}
	\end{minipage}
	\begin{minipage}[c]{0.32\linewidth} 
		\centering
		\includegraphics[width=1\textwidth]{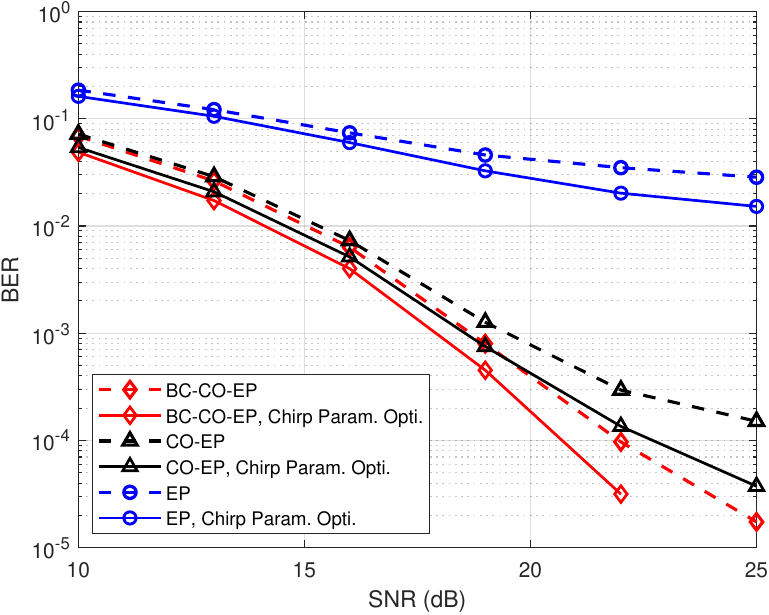}
		\caption{\textcolor{black}{Overall BER for three-user RSMA-AFDM system.}}
		\label{fig:three_user_overall}
	\end{minipage}
	\begin{minipage}[c]{0.32\linewidth} 
		\centering
		\includegraphics[width=1\textwidth]{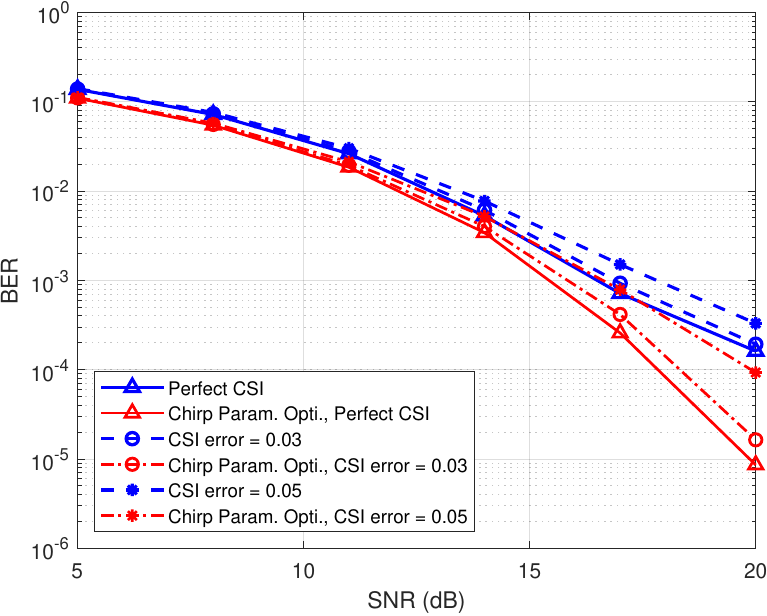}
		\caption{\textcolor{black}{Overall BER for belief consensus-based scheme under perfect and Imperfect CSI.}}
		\label{fig:csi_error}
	\end{minipage}
\end{figure*}
	
\begin{figure}
	\centering
	\includegraphics[width=0.34\textwidth]{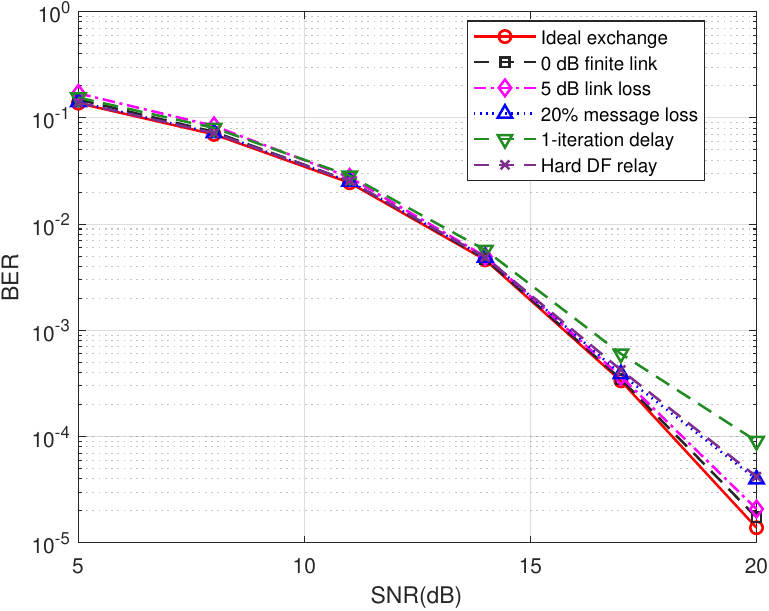}
	\caption{\textcolor{black}{Overall BER for belief consensus-based scheme under non-ideal collaboration links.}}
	\label{fig:link_impairments}
\end{figure}	
	
	\textcolor{black}{Fig. \ref{compcpode} compares the BER performance with and without the proposed PEP-guided $c_1$ optimization. CO-EP and BC-CO-EP represent the proposed decision-fusion-based and belief-consensus-based cooperative EP detectors, respectively. For all detection schemes, the optimized curves shift downward, confirming that minimizing the channel-averaged PEP improves the precoder-dependent cooperative effective distance. The two proposed cooperative EP detectors outperform the benchmark cooperative detectors over the entire SNR range. At a BER of \(10^{-4}\), the proposed decision-fusion-based and belief-consensus-based detectors achieve approximately \(3\,\mathrm{dB}\) and \(5\,\mathrm{dB}\) SNR gains, respectively, compared with cooperative CD-MAMP.}
	
	\textcolor{black}{Then, we evaluate the proposed schemes with $N_t = 8$ and $K=2$. We examine the BER of the common stream, the private streams, and the overall system, as shown in Fig. \ref{PNt8}. The results show that the PEP-guided chirp-parameter optimization improves the common-stream detection by selecting a precoder-dependent cooperative effective distance. In addition, both proposed cooperative EP detection schemes significantly outperform the non-cooperative EP detector, demonstrating the benefit of distributed information fusion.}

Here, we demonstrate the overall BER of a three-user RSMA-AFDM system. With $N_t = 4$, \(N=64\) and $l_{\mathrm{max}}=f_\mathrm{max}=2$, as shown in Fig. \ref{3userc1}, the proposed joint PEP-guided design selects the chirp indices \((m_1,m_2,m_3)=(7,8,16)\), yielding
\begin{equation}
(c_{1,1},c_{1,2},c_{1,3})
=\frac{[7,8,16]}{2N}
=[0.0546875,\;0.0625,\;0.125].
\end{equation} \textcolor{black}{As shown in Fig. \ref{fig:three_user_overall}, the optimized curves remain below their corresponding baselines for all three detectors, confirming the BER improvement obtained by the PEP-guided cooperative effective-distance optimization. In addition, both cooperative EP schemes significantly outperform the non-cooperative EP detector. These results demonstrate that the proposed $c_1$ optimization and cooperative detection framework can be extended to more users.}

To evaluate the robustness of the belief-consensus-based cooperative EP detector, imperfect CSI is modeled as \cite{9349154}
\begin{equation}
f_i=\hat f_i+\Delta f_i,\qquad
\tau_i=\hat\tau_i+\Delta\tau_i,\qquad
h_i=\hat h_i+\Delta h_i,
\end{equation}where \(\Delta(\cdot)\) denotes the CSI error and each perturbation is bounded by its corresponding error limit. Fig. \ref{fig:csi_error} compares the overall BER under perfect CSI and normalized error bounds of \(0.03\) and \(0.05\). The BER increases as the CSI error becomes larger, but all curves decrease monotonically with SNR and show no divergence or high-SNR error floor. This demonstrates that the proposed belief-consensus detector maintains stable convergence and reliable detection performance under bounded CSI errors.

Fig. \ref{fig:link_impairments} shows the robustness of the belief-consensus-based cooperative detector under non-ideal collaboration links. At low and moderate SNRs, all curves nearly overlap because the detection performance is mainly limited by the wireless-channel noise. As the SNR increases, ideal exchange achieves the lowest BER, while the 0-dB finite-link and 5-dB link-loss cases remain close to the ideal benchmark, indicating that the proposed detector is tolerant to moderate collaboration-link attenuation. Message loss and hard-decision DF relay introduce a larger performance degradation, whereas using one-iteration-old information causes the most pronounced loss. Nevertheless, all cases preserve a decreasing BER trend, demonstrating the robustness of the proposed belief-consensus detector against practical collaboration impairments.
	
\subsection{Complexity Analysis}
Finally, we evaluate the computational complexity of the proposed detectors. For offline chirp optimization, the search complexity is given by
$
	C_{\mathrm{search}}
	=N_{\mathrm{ch}}\left[M_0+(N_{\mathrm{cand}}+1)M_1\right]
	+N_{\mathrm{cand}}N^3,
$
where $N_{\mathrm{ch}}$ and $N_{\mathrm{cand}}$ denote the number of channel realizations and chirp candidates, respectively, while $M_0$ and $M_1$ represent the shared preprocessing and per-candidate mapping costs. The complexity of the discrete search is of limited practical concern, as the optimization is performed offline with sufficient computational time available.

For online detection, let $d=2N$ denote the dimension of the real-valued model in (\ref{ior}). Define
\begin{equation}
	\begin{gathered}
		F(d)=\frac{d^3-d}{6},\quad
		T(d)=\frac{d(d-1)}{2},\\
		\textcolor{black}{C_0=F(d)+(d+1)T(d)+d^3+d^2},\\
		\textcolor{black}{C_{\mathrm M}=F(d)+2T(d)},\qquad
		\textcolor{black}{C_{\mathrm E}=F(d)+2dT(d)+d^2}.
	\end{gathered}
\end{equation}
\textcolor{black}{Here, $C_0$ accounts for per-user whitening and formation of the local external likelihood parameters $(\mathbf J_v,\mathbf b_v)$, while $C_{\mathrm M}$ and $C_{\mathrm E}$ represent one MMSE and EP Gaussian update, respectively.} For CO-CD-MAMP, with $J_{\mathrm{MAMP}}$ iterations and constellation size $M_{\mathrm{mod}}$, the complexity is
\begin{equation}
	\begin{aligned}
		C_{\mathrm{CD}}^{\mathrm{core}}
		=&4K[5N^3+B(N)+N^2]
		+4J_{\mathrm{MAMP}}(5K+3)N^2\\
		&+NJ_{\mathrm{MAMP}}(J_{\mathrm{MAMP}}+1)
		+2J_{\mathrm{MAMP}}NM_{\mathrm{mod}},
	\end{aligned}
\end{equation}
where $B(N)=F(N)+(N+1)T(N)$.

Table~\ref{tab:detector_complexity} summarizes the arithmetic costs, using $N=64$, $K=2$, $L=4$, $J_{\mathrm{MAMP}}=5$, and QPSK ($M_{\mathrm{mod}}=4$). \textcolor{black}{The table reports multiplications for all $K$ receivers. The belief-consensus EP performs the same $L$ Gaussian updates per receiver as independent EP, but also aggregates $(\mathbf J_v,\mathbf b_v)$ once per block. Because this aggregation uses direct summation, it introduces no additional real multiplications relative to independent EP under the stated counting convention.} For the EP algorithm, its high complexity can also be reduced through certain methods \cite{11263921}.

\begin{table}[!htbp]
	\caption{\textcolor{black}{Reference real-multiplication subtotals per $K$-user common-stream block.}}
	\label{tab:detector_complexity}
	\centering
	\footnotesize
	\setlength{\tabcolsep}{3pt}
	\begin{tabular}{@{}lcr@{}}
		\hline
		Detector & Count & Example\\
		\hline
		MMSE & $K(C_0+C_{\mathrm M})$ & 7,754,624\\
		Decision-fusion MMSE & $K(C_0+C_{\mathrm M}+Kd)$ & 7,755,136\\
		Independent EP & $K(C_0+LC_{\mathrm E})$ & 26,596,352\\
		Decision-fusion EP & $K(C_0+LC_{\mathrm E}+Kd)$ & 26,596,864\\
		{Belief-consensus EP} & {$K(C_0+LC_{\mathrm E})$} & {26,596,352}\\
		CO-CD-MAMP (joint) & $C_{\mathrm{CD}}^{\mathrm{core}}$ & 12,985,728\\
		\hline
	\end{tabular}
\end{table}

\color{black}	
	\section{Conclusion}
\textcolor{black}{This paper has investigated cooperative RSMA-AFDM transmission with
user-dependent chirp design for reliable common-stream detection.
A conditional PEP expression was derived with the cooperative
precoder explicitly included in the effective-distance term, thereby
establishing how the DAFT-domain channel representation influences
the transmitted common-stream direction and its detection reliability.
Based on this analysis, a PEP-guided $c_1$ optimization framework was
formulated for multiuser cooperative RSMA-AFDM systems. The design
adapts the chirp parameters of different users to minimize the
channel-averaged cooperative PEP, without requiring inter-user
DAFT-domain path-position separation. Two distributed EP-based
cooperative detection schemes, based on decision fusion and belief
consensus, were also developed. Numerical results demonstrate the performance benefits of the proposed chirp
design and cooperative detection methods. The general PEP formulation
and chirp-optimization framework provide a basis for extension to
systems with more cooperating users.}

	\bibliographystyle{IEEEtran}
	\bibliography{AFDM}
\end{document}